\documentclass[aps,prb,superscriptaddress,twocolumn]{revtex4-2}
\usepackage[cp1251]{inputenc}
\usepackage{graphicx}
\usepackage{float}
\usepackage{tabularx}
\usepackage{booktabs}
\usepackage{textcomp}
\usepackage{multirow}
\usepackage{placeins}
\usepackage{amsmath}
\usepackage{amsfonts}
\usepackage{graphicx}
\usepackage{graphics}
\usepackage{amssymb}
\usepackage{amsmath}
\usepackage{color}
\usepackage[normalem]{ulem}

\newcommand {\cb} {\color{black}}

\newcommand{\beq}{\begin{equation}}
	\newcommand{\eeq}{\end{equation}}
\newcommand{\bea}{\begin{eqnarray}}
	\newcommand{\eea}{\end{eqnarray}}

\usepackage{hyperref}  
\begin{document}
\title{Emergence of Ferromagnetism from Planar Defects in EuSn$_2$As$_2$ Antiferromagnet
}
\author{A. Yu. Levakhova}
\email[E-mail:]{levakhovaayu@lebedev.ru}
\affiliation{V.L. Ginzburg Research Centre for High-Temperature Superconductivity and Quantum Materials, P.N. Lebedev Physical Institute, RAS, Moscow 119991, Russia}
\author{A. L. Vasiliev}
\affiliation{National Research Centre``Kurchatov Institute'', Moscow 123182, Russia}
\affiliation{Moscow Institute of Physics and Technology, Dolgoprudny,
	Moscow district, 141701, Russia}
\author{N. S. Pavlov}
\affiliation{Institute for Electrophysics, RAS, Ekaterinburg, 620016, Russia}
\author{A. V. Ovcharov}
\author{V. I. Bondarenko}
\affiliation{National Research Centre``Kurchatov Institute'', Moscow 123182, Russia}
\author{A. V. Sadakov}
\author{K. S. Pervakov}
\author{V. A. Vlasenko}
\affiliation{V.L. Ginzburg Research Centre for High-Temperature Superconductivity and Quantum Materials, P.N. Lebedev Physical Institute, RAS, Moscow 119991, Russia}
\author{V. M. Pudalov}
\affiliation{V.L. Ginzburg Research Centre for High-Temperature Superconductivity and Quantum Materials, P.N. Lebedev Physical Institute of RAS, Moscow 119991, Russia}
\affiliation{National Research University ``Higher school of economics'', 101001, Moscow}

\date{\today{}}

\begin{abstract}
We  report  an observation and  study of monolayer thick 
structural defects in the antiferromagnetic (AFM) layered semimetal  EuSn$_2$As$_2$.
The high resolution transmission electron microscopy revealed the presence of planar defects in the lattice of the studied single crystals. Using a combination of microstructural and DFT analysis we  demonstrated that a single planar defect forms a monolayer of a distinct phase EuSnAs$_2$,  that is different from the  EuSn$_2$As$_2$  phase of the bulk lattice. 
The  smaller distance between Eu layers in the planar defect  as well as odd number of Eu atoms in the formula of the {\cb defect (EuSnAs$_2$) and of the defect supercell} (Eu$_7$Sn$_{12}$As$_{14}$) 
promote formation of  local ferromagnetic (FM) ordering of the Eu atoms. 
 This suggests that on average, the planar defects form a 
 {\cb  ferrimagnetic} minority   phase in the antiferromagnetic (AFM) host lattice.  The obtained results   may explain  several puzzling features in magnetic properties of various A-type AFM materials, such as  the nonlinear magnetization and ferromagnetic-type hysteresis in low in-plane fields,  the  upturn of the magnetic susceptibility in the AFM state at temperatures approaching zero,  and the emergence of the ferromagnetic-type resonance line in the electron spin resonance spectrum of the AFM state. 
  {\cb The present finding provides new insight into understanding the mechanism of ferromagnetism 
 in quasi-two-dimensional AFM materials: it offers a new way to create an  AFM-FM interface by self-docking 
 two different crystalline phases of a chemically pure compound.  
 Thus, this result will facilitate the design and  implementation  of natural AFM-FM interfaces.}
\end{abstract}

\keywords{Layered semimetals, antiferromagnets,  electron microscopy, planar defects, ferromagnetic defects, HRTEM, DFT }
\maketitle

\section{Introduction}
\label{introduction}
In recent years, great attention has been attracted by layered compounds which comprise  weakly bonded
layers of high spin atoms (Eu, Gd, Mn, Co, Cr) and charge reservoir layers of metal arsenides, phosphides, tellurides \cite{chang_Science_2013, otrokov_Nature_2019, zhang_PRL_2019, jo_PRB_2020, li_PRB_2019}. 
	These nonsuperconducting  semimetal compounds  offer  fascinating physics 
  due to propagation of charged carriers in alternating magnetic field of the antiferromagnetically (AFM) ordered host lattice. Both,  topologically non-trivial (MnBi$_2$Te$_4$, EuIn$_2$As$_2$, EuCd$_2$As$_2$, etc.) 
\cite{zhang_PRL_2019, otrokov_Nature_2019, li_PRB_2019, li_PRX_2019, li_PRB_2021, xu_PRL_2019, pierantozzi_PNAS_2022} 
 and trivial (EuSn$_2$As$_2$, EuFe$_2$As$_2$,  EuSn$_2$P$_2$,  EuMg$_2$Bi$_2$, etc.) 
 \cite{arguilla_InChemFront_2017, pakhira_PRB_2021, golov_JMMM_2022, golov_PRB_2022, chen_ChPhysLet_2020, li_PRX_2019, li_PRB_2021, sanchez_PRB_2021, grigorev-CommunMater_2025} materials are in the focus of research interest.  
 	No less interesting is the physics of coexistence in arsenides of superconducting pairing with a non-trivial topology
 	  \cite{morgun-PRB_2019}  or  with AFM ordering  of high-spin atoms (Eu, Gd, etc.) \cite{kim_PRB_2021, khlybov_2009}.

The quasi-two-dimensional (Q2D) materials 
provide a unique platform for designing spintronic devices down to the atomic thin scale \cite{yang_PRX(2024)}.
However, this prospect is hampered by the limited number of antiferromagnetic Q2D materials that could be spatially coupled with ferromagnetic (FM) materials, as is necessary for the engineering of various  device applications. 
If the Q2D antiferromagnetism could be converted to Q2D ferromagnetism, 
the range of 2D magnetic structures and their potential applications would be 
significantly broadened. {\cb Creating two-dimensional ferromagnetism from van der Waals antiferromagnets 
would inherently combine the advantages of both ferromagnets and antiferromagnets 
(e.g., the  robust magnetic state of ferromagnets and the ultrafast switching of antiferromagnets), 
which could lead to potential breakthroughs in high-speed, low-power spintronics.} 

A tightly related issue is the physics of the AFM-FM interface.  
For realizing such interface, {\cb commonly, artificial epitaxial  or  assembled bilayers  \cite{gong_NatCom(2023)}  
of the exchange coupled FM and AFM layers are used.}
The presence of disorder at the artificial interfaces hinders the native spin arrangements and exchange interaction control  
between the FM and AFM phases. 
In this  regard, the natural FM-AFM interfaces which are formed inside the bulk crystals
represent a unique platform for studying the intrinsic physics of the FM-AFM interface 
 {\cb and exploring   	the mechanism of ferromagnetism in quasi-two-dimensional AFM materials 
	which is a critical challenge for realizing or engineering natural AFM-FM interfaces. This paper presents evidence that the polymorphic modification 
	EuSn$^{(4+)}$As$_2$ (which emerges in the host EuSn$^{(2+)}_2$As$_2$ phase during high-temperature synthesis) forms a natural interface between antiferromagnetic and ferromagnetic regions, owing to the mixed valence of the Sn atoms.
We investigated the lattice microstructure, electron band structure, and magnetic properties of the resulting natural  stoihiometric structure containing FM-inclusions in the AFM matrix. }

Our study focuses on a typical AFM Q2D semimetal, 
EuSn$_2$As$_2$ (ESA), because  beside  conventional  properties intrinsic to A-type Q2D 
antiferromagnet \cite{arguilla_InChemFront_2017, golov_JMMM_2022},
signatures of ferromagnetism have been noticed in several preceding studies.
 For instance, in DC-magnetization  a weak nonlinearity 
 was observed in low fields, $< 0.03$\,T, when field is applied in the easy magnetization $ab$-plane \cite{li_PRB_2021, chen_ChPhysLet_2020}. The AC-susceptibility below the Neel temperature $T_N$  
often shows an upturn towards lowest temperatures in the same low field range \cite{pakhira_PRB_2021, li_PRB_2021, li_PRX_2019, chen_ChPhysLet_2020}. 
Furthermore,  in electron spin resonance measurements \cite{golov_JMMM_2022}, in the AFM state of ESA,
an additional resonance  was observed, atypical for the AFM state but quite consistent with its interpretation as ferromagnetic resonance \cite{golov_JMMM_2022, talanov_tbp}. 
All these features hint at the potential presence of a burried ferromagnetism {\cb  (or ferrimagnetism)}  in the bulk AFM crystal.

Although these FM-like features were observed 
before, to our knowledge, their microscopic origin has not been understood and explained until now. 

In order to  elucidate these puzzling features, and reveal the nature of the 
{\cb ferromagnetic-type} response in a layered AFM crystal we performed and report here detailed microstructural investigation of several EuSn$_2$As$_2$ single crystals. The main outcome of our research is the detection  and identification of about a monolayer thick planar defects  in the layered host crystal. The  defects are randomly distributed in the bulk with a total volume  concentration of  $\sim 3\%$.

We have identified  local composition of the planar defects and found that the {\cb monolayer of the defect} has the local composition of EuSnAs$_2$,  with Sn atoms in the (+4) oxidation state. In contrast to a regular host lattice,
the {\cb EuSnAs$_2$ monolayer}  has an extra row of Eu atoms and missing half of Sn-atoms;  nevertheless, it remains stoichiometric due to the  valence skipping of Sn atoms.  This compound has a cubic crystal structure which is  joined  along the (111) plane with the parent trigonal structure. 
{\cb The transition from the regular ESA lattice to the EuSnAs$_2$ monolayer is mediated by a transition layer (``supercell'') consisting of 33 atoms.
The supercell  incorporates the EuSnAs$_2$ monolayer}, and has a chemical formula Eu$_7$Sn$_{12}$As$_{14}$. {\cb Both  the defect monolayer and the  supercell contain an odd number of Eu atoms (1 and 7, respectively)} and, correspondingly, 
are anticipated to have  a non-zero  magnetic moment. {\cb In this aspect, the supercell is similar to a ferrimagnet, whereas the monolayer defect - to ferromagnet.}

We also performed precise magnetization measurements in weak external fields ($<0.05$T) and revealed  the FM-type hysteresis when the external field is applied in the easy $ab$ plane. The hysteresis disappears as temperature raises above 24\,K, close to the Neel temperature $T_N$ of the host lattice, demonstrating  an intimate relationship between the FM- and AFM-type ordering in the crystal. The FM-type hysteresis explains the magnetization nonlinearity that is  often observed in experiments in weak fields.

Furthermore, we performed DFT  calculations of the band structure and magnetization of the defect. Their results confirmed that 
the monolyer planar defect  (EuSnAs$_2$) and, hence, the  supercell  (Ei$_7$Sn$_{12}$As$_{14}$)  posess a  {\cb magnetic moment  almost equal to  one Eu atom moment, $6.77\mu_B$ per f.u. } 
 Thus, the crystal lattice of the layered A-type   antiferromagnet, ESA,
comprises a sample-dependent amount of defects with a nonzero magnetic moments which experience ordering at low temperatures $T \leq T_N$, in the AFM state of the host crystal. 

The existence  of the ferromagnetically polarized   monolayer-thick 
	planar defects imbedded into the AFM matrix may explain the above mentioned features in the AFM state of the host lattice,  and hence suggests solution of the long-standing puzzle of the FM-like responce of the layered vdW crystals in the AFM state.
Thus, our results  unveil the origin of such puzzling features as magnetization nonlinearity and hysteresis, susceptibility  upturn with cooling below $T_N$, and splitting of the ESR resonance. We show that all these features originate from a weak ferromagnetic magnetization of  monolayer-thick planar defects randomly distributed in the bulk crystal, {\cb whereas the defect supercells  and the real bulk ESA crystal may be viewed as  ferrimagnets}.  
{\cb The present findings  provide new insights into the interface structure and propose a new approach to designing stoichiometric FM-AFM interfaces based on layered van der Waals crystals.}
It is worthnoting, the emergence  of ferromagnetism in microscopic areas of semiconductor and its coesistence with the host AFM  lattice  was theoretically considered in 1968 by Eduard Nagaev  \cite{nagaev_JETP_1968}.

{\bf This paper is organized as follows:}
in Section \ref{sec:methods} we  describe  crystal lattice and magnetic structure of ESA. We also describe the samples under study, their synthesis, methods of magnetic and microstructural measurements. In section \ref{sec:chi_&_M}  we present our results of AC and DC magnetic measurements which manifest ferromagnetic-type features. In Section \ref{sec:TEM}  we present and analyse  the main experimental data   of the transmission electron microscopy where we revealed  defects,  identified the chemical formula of their composition, and their local lattice structure. In section \ref{sec:DFT} 
we show  the results of the numerical DFT  calculations of the band structure and  magnetization for the defect supercell. 
 Finally, we provide a brief summary and discussion of the results in section 
 \ref{sec:discussion} and \ref{sec:conclusions}. More detailed information is presented in Supplementary materials (SM) \cite{SM}.

\section{Samples and Methods}
\label{sec:methods}
\subsection{The studied compound}
\begin{figure}
	\includegraphics[width=220pt]{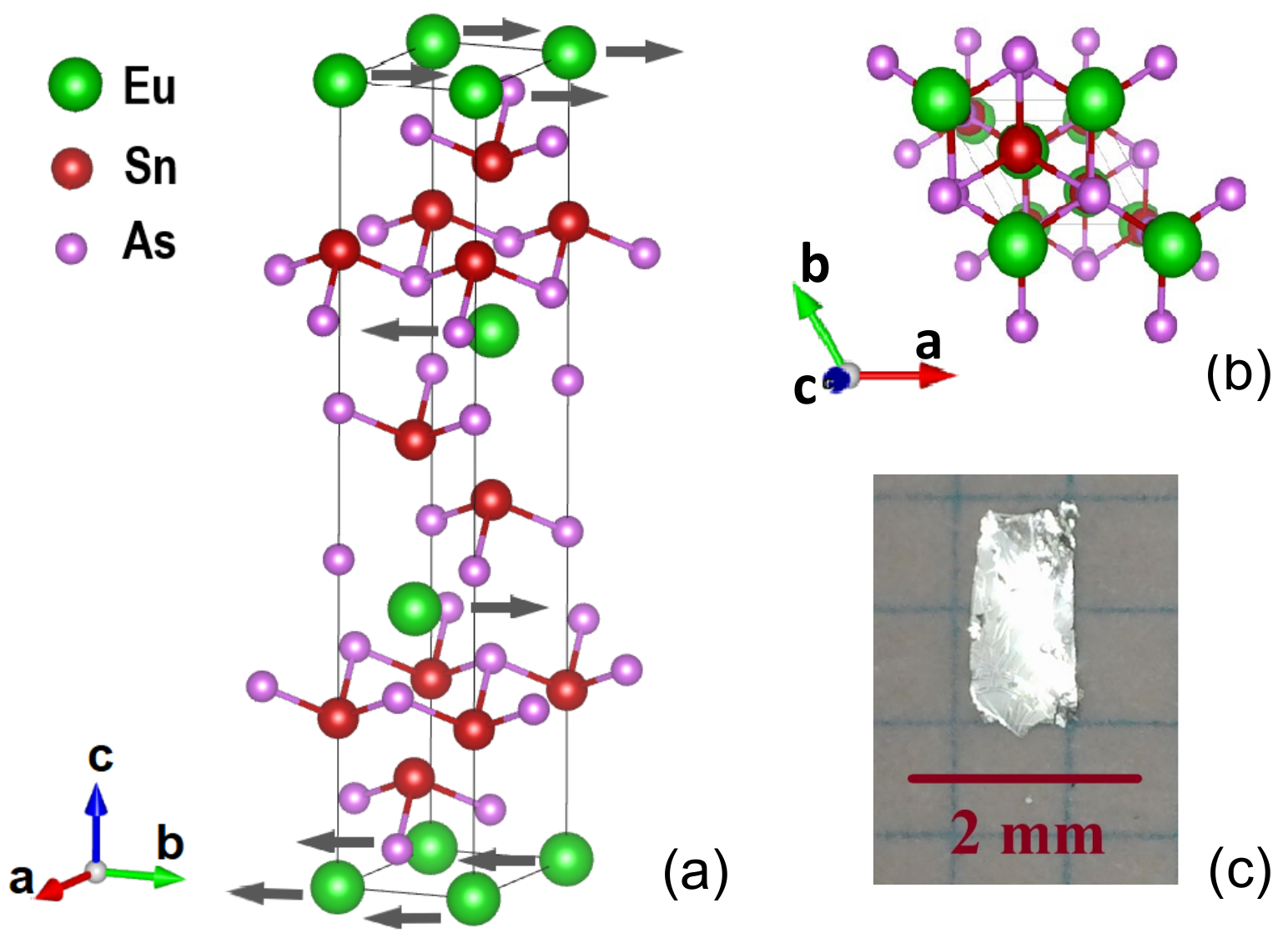}
	\caption{(a,b)  Crystal lattice and magnetic structure of EuSn$_2$As$_2$ in the two projections. The
		crystal structure of EuSn$_2$As$_2$ is rhombohedral with the space group $R\bar{3}\,m$ \cite{arguilla_InChemFront_2017}. The Eu spin sub-lattice is of A-type antiferromagnetic state with the $ab$ 
		easy magnetization plane (adapted from Ref.~\cite{golov_JMMM_2022}). Black arrows show Eu atoms magnetization directon. (c) Bulk sample view.
	} 
	\label{fig:1}
\end{figure}

 EuSn$_2$As$_2$,   a representative layered  semimetal  was first synthesized 
by M. Q. Arguilla, et al. \cite{arguilla_InChemFront_2017}. Its lattice consists of Eu-layers in the $ab$ plane alternating along the $c$ axis with pair of SnAs layers (Fig.~\ref{fig:1}). 
{\cb Due to rather small distance between Eu atoms within the layer,
the ferromagnetic-type intralayer exchange interaction prevails  \cite{sun_SciChina_2021}}
and the large magnetic moments of Eu atoms align ferromagneticaly (FM) along  the Eu layers (in the $ab$ plane). 
Upon cooling below $T_N\approx 24$K, the Eu layers align antiferromagnetically (AFM) and form an A-type AFM structure \cite{pakhira_PRB_2021, pervakov_PRB_2026}, where the magnetization vector remains within  $ab$ planes and rotates by $\pi$ from layer to layer (Fig.~\ref{fig:1}). As a result, the ideal  EuSn$_2$As$_2$ crystals exhibit text-book magnetic properties, conventional for easy-plane antiferromagnets:
a linear magnetization field dependence extending up to the field of complete spin polarization, and a growth of the spin susceptibility  in the paramagnetic state with cooling down to $T_N$
\cite{pakhira_PRB_2021}.

\subsection{Synthesis of  EuSn$_2$As$_2$ crystals and samples characterization}

The high-purity materials used for  growing  EuSn$_2$As$_2$ single crystals were pieces of Eu (99.95\%, LANHIT), Sn (99.99\%, LANHIT) and As (99.9999\%, LANHIT). The SnAs precursor was syntesized beforehand, as described in Ref. ~\cite{SnAs-PRB_2019},
from pure elements, Sn and As. 
The sealed quartz ampoule with elements taken in a stoichiometric ratio of 1 : 1 was slowly heated to
 600\,$^\circ$C, held for 48\,h to form the phase, then heated to 800\,$^\circ$C and held for 24\,h to homogenize the melt. 
 Next, the EuSn$_2$As$_2$ single crystals  were grown by self-flux technique as described in the  \cite{golov_JMMM_2022}.
The SnAs compound and elemental Eu (99.95\%) pieces were thoroughly mixed in  molar ratio (4:1)
to ensure the synthesis reaction and self-flux growth in a single process:
$$
\textrm{Eu +2SnAs + flux (2SnAs)}  \rightarrow \textrm{EuSn}_2\textrm{As}_2 + \textrm{flux (2SnAs)}.
$$
The prepared mixture pressed into a pellet was loaded  into an alumina crucible, 
placed in a quartz ampoule which was sealed in Ar atmosphere under excess pressure of 0.2\,atm. The sample was heated in a furnace for 9 hours up to  850\,$^\circ$C, then held at this temperature for 24 hours, and cooled  down at a rate of  2\,$^\circ$C/h down to 500\,$^\circ$C, 
and then held for 24 hours. The furnace was then turned off and the ampoule, together with the furnace, was cooled in air. The synthesized aggregate was separated by mechanical cleaving, and crystals with dimensions about $2\times 1\times 0.2$\,mm were selected for the measurements (Fig.~\ref{fig:1}c).

Several small single crystals of EuSn$_2$As$_2$ with a mirror surface in the $ab$ plane have been selected 
for structural studies  (see Fig.~\ref{fig:1}c and also Fig.~1a in \cite{SM}).  Phase composition  was tested using Rigaku MiniFlex 600 powder diffractometer.  Lattice structure of the grown bulk crystals was characterized  using  X-ray crystal diffractometers  (X'PertPro-MRD and Tongda TD-5000).  The X-ray diffraction (XRD) measurements didn't reveal any structural defects 
except for   minor 	random misorientation (twisting) of crystal layers in the $ab$-plane (see SM \cite{SM}). 
 XRD  measurements of the grown crystals confirmed previously  reported  lattice parameters $a=b\approx 4.2113(9)$\AA, $c=26.329(5)$\AA\,  and the elementary cell volume $V_c=404.38(19)$\AA$^3$ (for more detail, see SM  \cite{SM}).
 Elemental composition on the microscale was determined  from EDX analysis using JEOL JSM 7001F scanning electron microscope equipped with INCA X-act analyzer. Additionally,  lattice site occupancies  were determined from high precision crystal XRD, using  Tongda TD-5000 crystal diffractometer (see Section IV SM  \cite{SM}).

\subsection{Methods and techniques of the crystal nanostructural investigation}

Crystal structure and elemental composition on the nanoscale   was studied with 
	a transmission/scanning electron microscope (TEM/STEM) Titan Themis Z (Thermo Fisher Scientific, USA) at an accelerating voltage of 200 kV. The microscope equipped with a probe spherical aberration corrector (the spatial resolution better than 0.5\AA), high-angle annular dark-field (HAADF) detector (Fischione, USA), electron energy loss spectrometer (EELS) (GIF, Gatan, USA) and energy dispersive X-ray spectrometer (EDXS) (EDAX, USA). 

 Samples (lamellae)  for TEM/STEM investigation were prepared by lift-out focused ion beam (FIB) technique in the dual beam scanning electron-ion microscope Helios Nanolab 660i (Thermo Fisher Scientific, USA) equipped with Ga$^+$ focused ion beam source and a micro-manipulator (Omniprobe). The lamellae  had lateral dimensions of $\sim 15\times 5\mu$m$^2$ (see Fig.~1b of \cite{SM}) in the  plane containing the $c$-axis and thickness of  less than $\approx 15$\,nm. 

 The high resolution (HR) HAADF STEM image simulations were performed with Dr.~Probe software 
 for  several lamellae of different thicknesses.
Digital Micrograph (Gatan, USA), TIA (FEI, USA) software was used for image processing. 

\subsection{Band structure computation}
\label{sec:BS calcuation}
The electronic structure of EuSn$_2$As$_2$ and Eu$_7$Sn$_{12}$As$_{14}$ was calculated in the framework of DFT \cite{mundy_APL_2014} using the generalized gradient approximation DFT/GGA implemented in the WIEN2k software package \cite{blaha_2020}  (full-potential linearized augmented plane wave (FP-LAPW) method). 
The generalized gradient approximation (GGA) in the form of the Perdew-Burke-Ernzerhof (PBE) exchange-correlation functional~\cite{DFT_PBE} was employed, and the spin-orbit interaction was also included. The local Coulomb interaction on the Eu-4f state was taken into account in the framework of DFT+U \cite{DFT_U} with values $U=6.8$\,eV and $J=0.7$\,eV. The correction for the double counting was taken in the fully localized limit form (FLL)  \cite{FLLpaper1}. 

\section{AC-magnetic susceptibility and DC-magnetization measurements}
\label{sec:chi_&_M}
\begin{figure}
		\includegraphics[width=210pt]{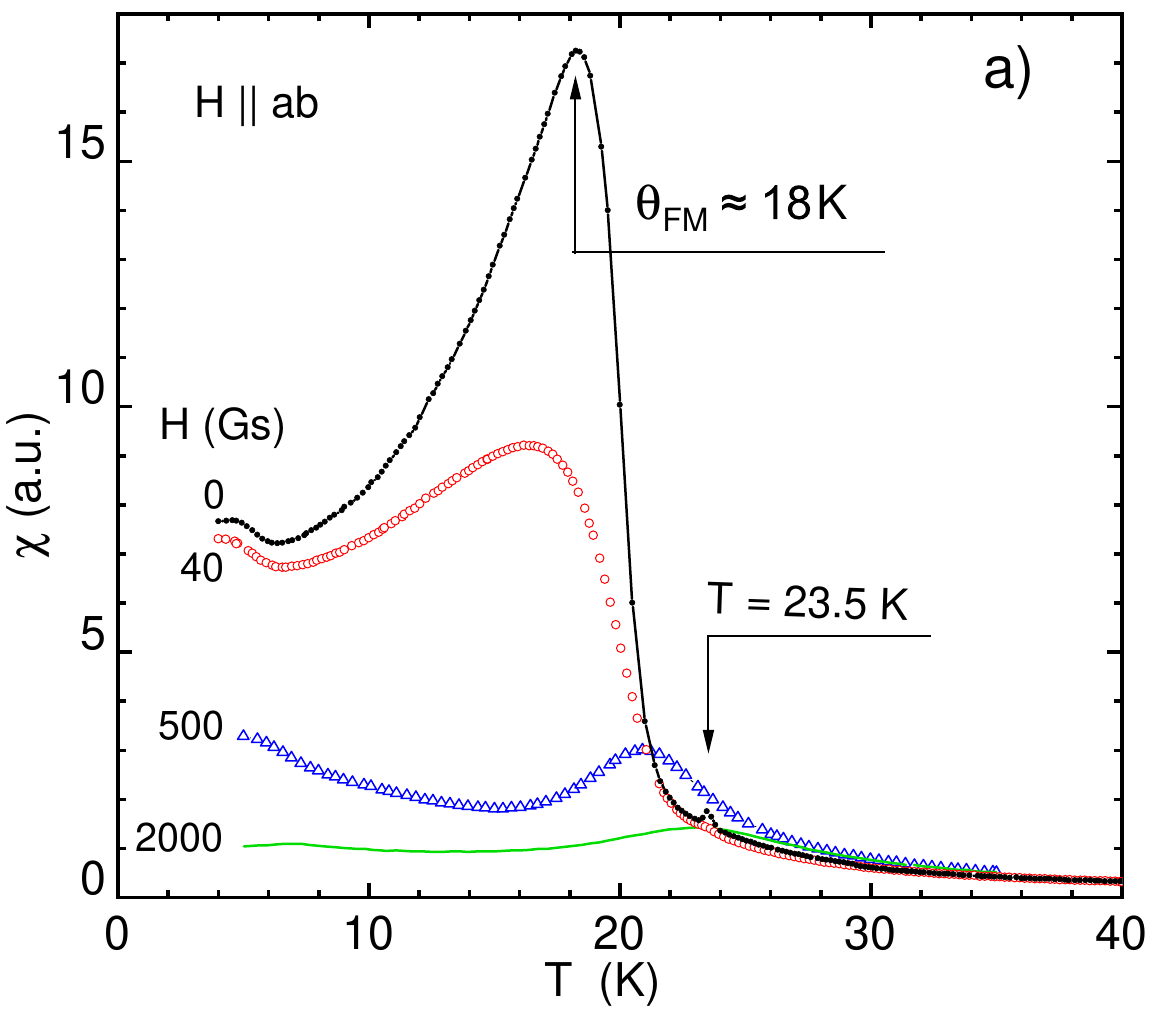}
	  \includegraphics[width=220pt]{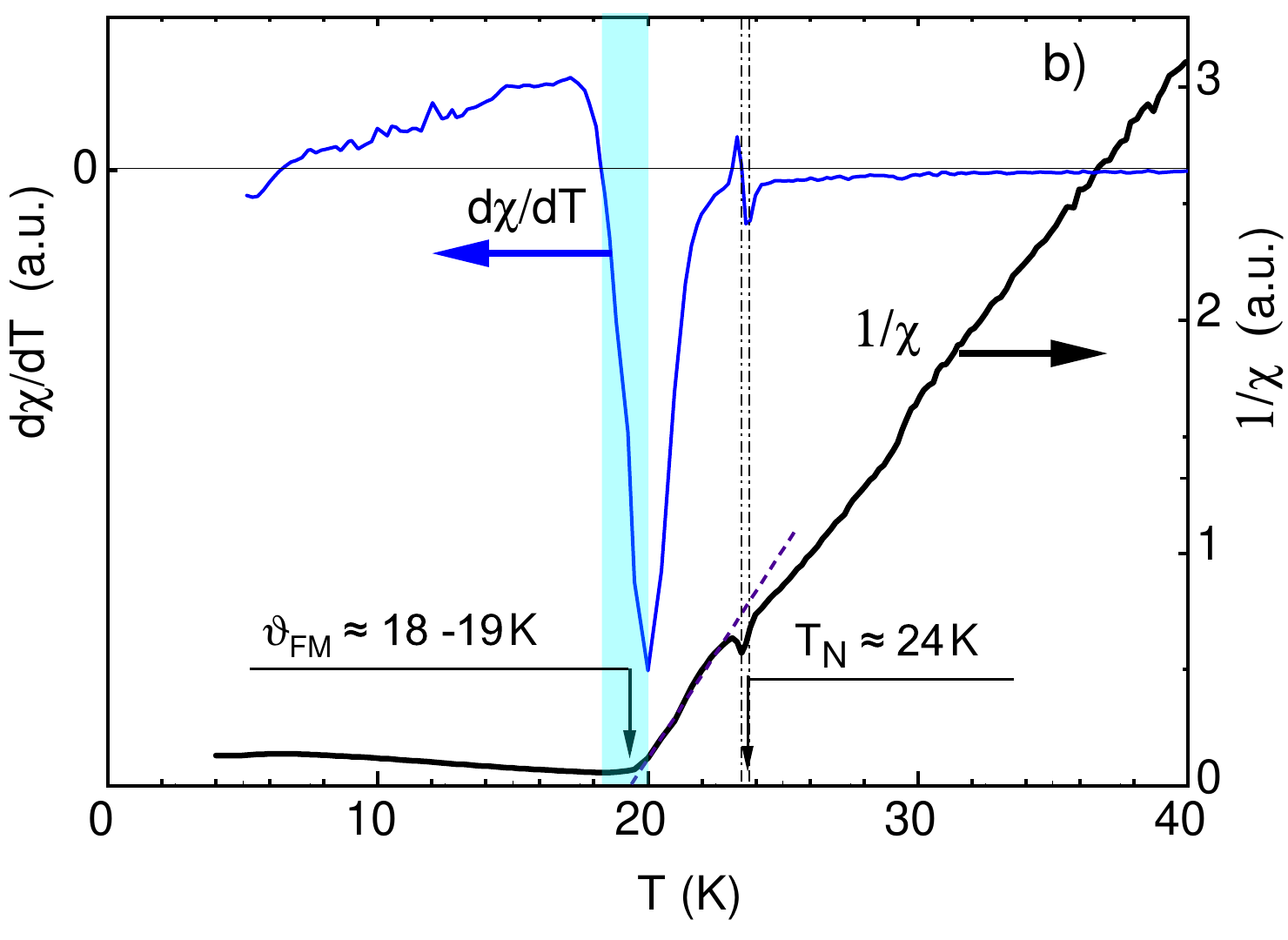}
	\caption{{\cb (a) Temperature dependence of the AC-magnetic susceptibility   at various  DC fields applied in the easy $ab$ plane; (b) temperature dependences of the inverse susceptibility $1/\chi$  and the derivative $d\chi/dT$ (both at $H=0$) for EuSn$_2$As$_2$ single crystal with planar defects.  
	 Vertical arrows mark  the Neel temperature $T_N$, and an approximate temperature of the ferromagnetic defects ordering $\Theta_{\rm FM}$. 
	Dashed line depicts a Curie-type linear dependence,  used to estimate the $\Theta_{\rm FM}$ value.
	The hatched blue strip depicts  the interval of ferromagnetic ordering. }
	}
	\label{fig:chi(T)}
\end{figure}

Temperature dependence of the AC-magnetic susceptibility and DC magnetization 
were measured using MPMS-7 SQUID magnetometer, in the wide range of temperatures, from the antiferromagnetic  to the paramagnetic state.
Temperature dependence of the AC-susceptibility (Fig.~\ref{fig:chi(T)}a), is typical for EuSn$_2$As$_2$ compound at zero applied magnetic field. In the paramagnetic state, as temperature decreases to $T_N$, the  AC magnetic susceptibility grows in the conventional manner  manifesting the dominating FM-type interaction between Eu spins \cite{pakhira_PRB_2021}.

For the  ideal easy-plane antiferromagnet  ESA,
within the molecular field theory, the out of plane susceptibility  component $\chi_c$   is expected to remain constant below $T_N$  down  to $T=0$, and the in-plane component $\chi_{ab}$ - to decrease by a factor of 2 \cite{pakhira_PRB_2021}.
However, in experiments the  $\chi_{ab}$ susceptibility component (in the easy-plane) often shows again a Curie-like upturn below $T_N$, as we mentioned above. This sample-dependent  growth of the AC susceptibility with cooling in the  AFM state was observed in Refs.~\cite{chen_ChPhysLet_2020, li_PRB_2021, pakhira_PRB_2021}, and in  many other works; it was conjectured to be due to anysotropic  magnetic impurities \cite{pakhira_PRB_2021}, with  no proper experimental and theoretical verification.

As an exaggerated illustration, we show in Fig.~\ref{fig:chi(T)}a the data for the sample where this upturn is even much larger than the conventional peak at $T_N$.  In a field of $0.01$\,T, the upturn weakens  and at $0.1-0.2$\,T is completely suppressed. It is worthnoting, such fields are much smaller than the field of complete Eu-spin polarization  in ESA
($H_{\rm s}\approx 3.6$\,T) \cite{pakhira_PRB_2021, golov_JMMM_2022,  pervakov_PRB_2026}. The growth  of $\chi$ at  $T<T_N$  (Fig.~\ref{fig:chi(T)}) manifests an apparent FM-type  transition  at $\Theta_{\rm FM} \approx 18 - 19$\,K. 
{\cb Because of a complex superposition of the AFM- and FM-contributions, the antiferromagnetic ordering temperature 
	$T_N\approx 24$\,K  was determined from the temperature dependence of the magnetization $M(H)$ curves and, particularly, the saturation field $H_s$  (see Fig.~9 of SM \cite{SM}), where the AFM transition is most clearly  pronounced. }

\begin{figure}
	\includegraphics[width=210pt]{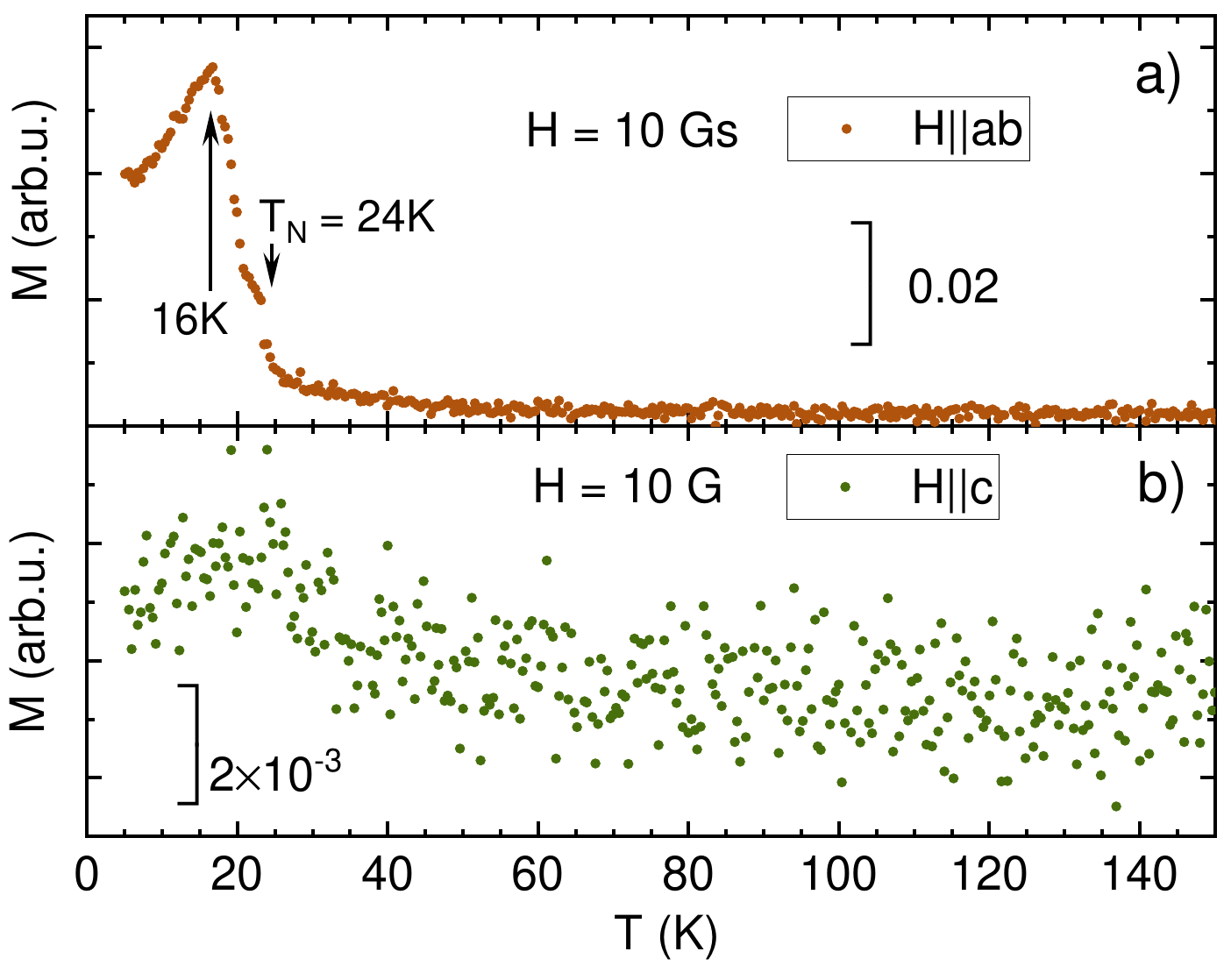}
	\caption{{\cb 
		Temperature dependence of the low-field magnetization for two field orientations: a) $H\| (ab)$ and b) $H\| c$.
Measurements were done with a fixed DC field $H=10$\,Gauss. Vertical brackets show approximate scale in emu/g.}
	}
	\label{fig:M(T)}
\end{figure}

{\cb  To test the hypothesis that the weak ferromagnetic-type  response  occurring below  19\,K is anisotropic in nature and appears only when a magnetic field is applied in the $ab$ plane, we measured the magnetization, using SQUID magnetometer, on the smallest available sample weighing only $\sim 50$ micrograms. However, since the magnetization signal from the sample in low fields was close to the sensitivity limit of the SQUID magnetometer, we were able to measure the temperature dependence of the magnetic moment, but we could not directly measure the susceptibility using the AC method. Although we succeeded in recording the temperature dependencies and confirming the anisotropy of the effect, 	the uncertainty in determining the sample mass was on the order of 20 micrograms;
 therefore, it is impossible to reliably determine absolute values of magnetization, so we present the dependences in Figs.~\ref{fig:M(T)} in relative units.}

{\cb	
For $H\|(ab)$-plane (Fig.~\ref{fig:M(T)}a) the in-plane magnetization exhibits an upturn  below $T_N\approx 24$K,  consistent with AC-susceptibility upturn in Fig.~\ref{fig:chi(T)}. In contrast, when magnetic field is applied in perpendicular direction $H\|c$, deviation of $M(T)$ from the horizontal line is almost within experimental uncertainty (see Fig.~\ref{fig:M(T)}b).  
 The minor increase  in $M(T)$  for $H\|c$ in Fig.~\ref{fig:M(T)}b is 20 times smaller than for $H\|(ab)$ in  Fig.~\ref{fig:M(T)}a. It is observed in the interval 25-50K and may be associated with the conventional effect of developing AFM 
 fluctuations in the vicinity of magnetic transition.
The drastically different $M(T)$ dependences  in Figs.~\ref{fig:M(T)}a and \ref{fig:M(T)}b demonstrate that the magnetic moments responding below $T_N$ to a weak probing field of 1\,mT  lie exclusively in the $(ab)$ plane. }

\begin{figure*}[ht]
		\includegraphics[width=162pt]{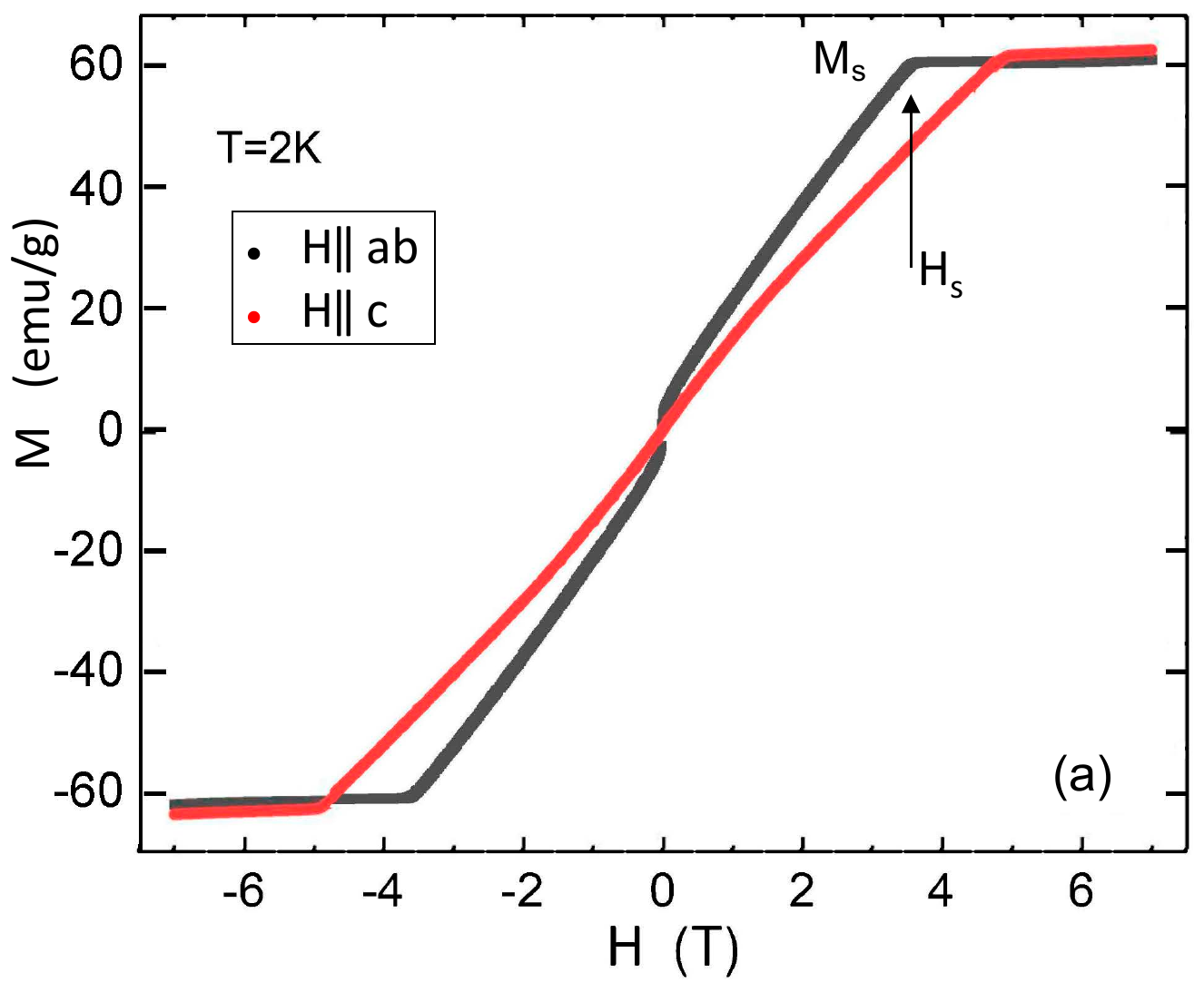}
		\includegraphics[width=173pt]{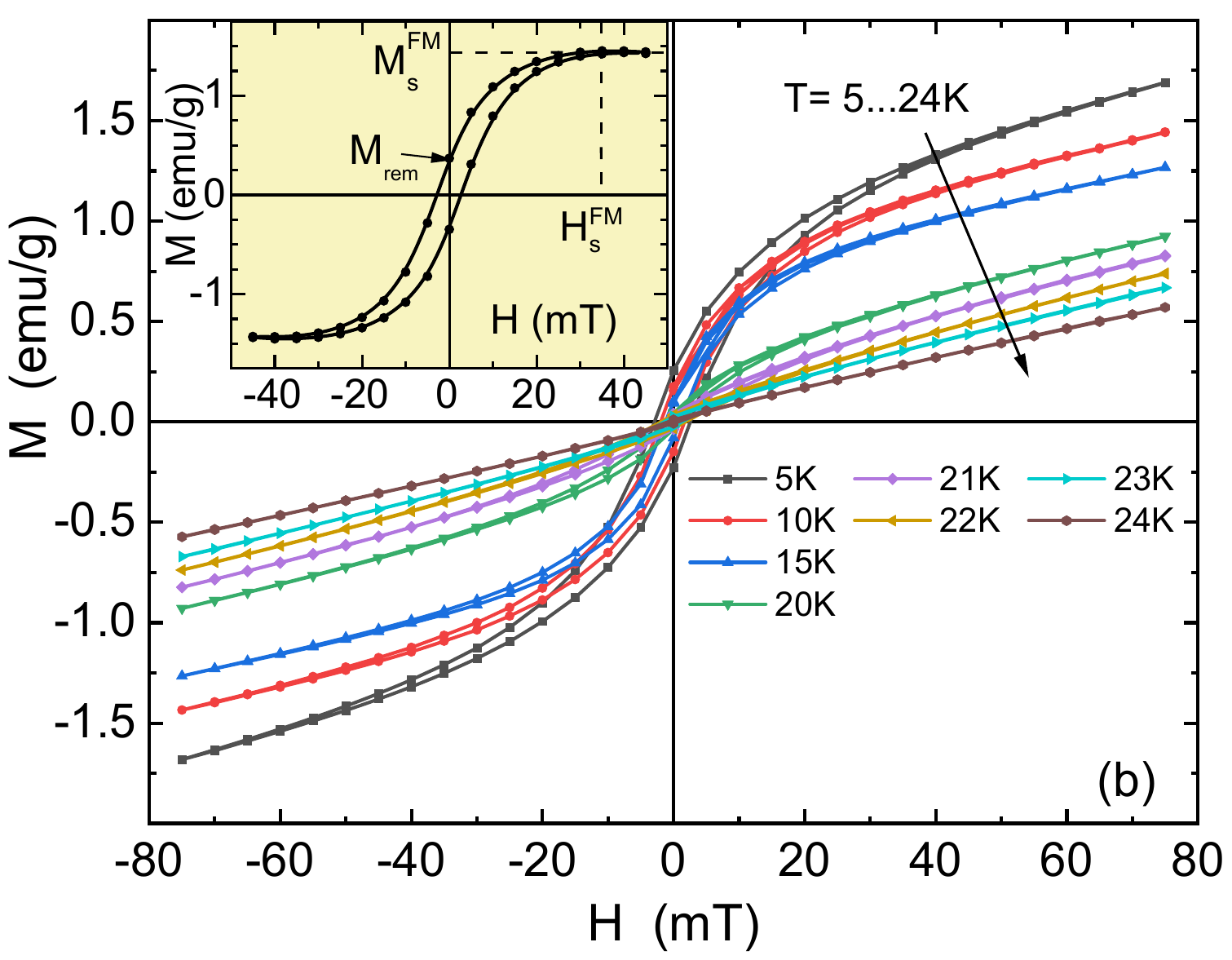}
		\includegraphics[width=162pt]{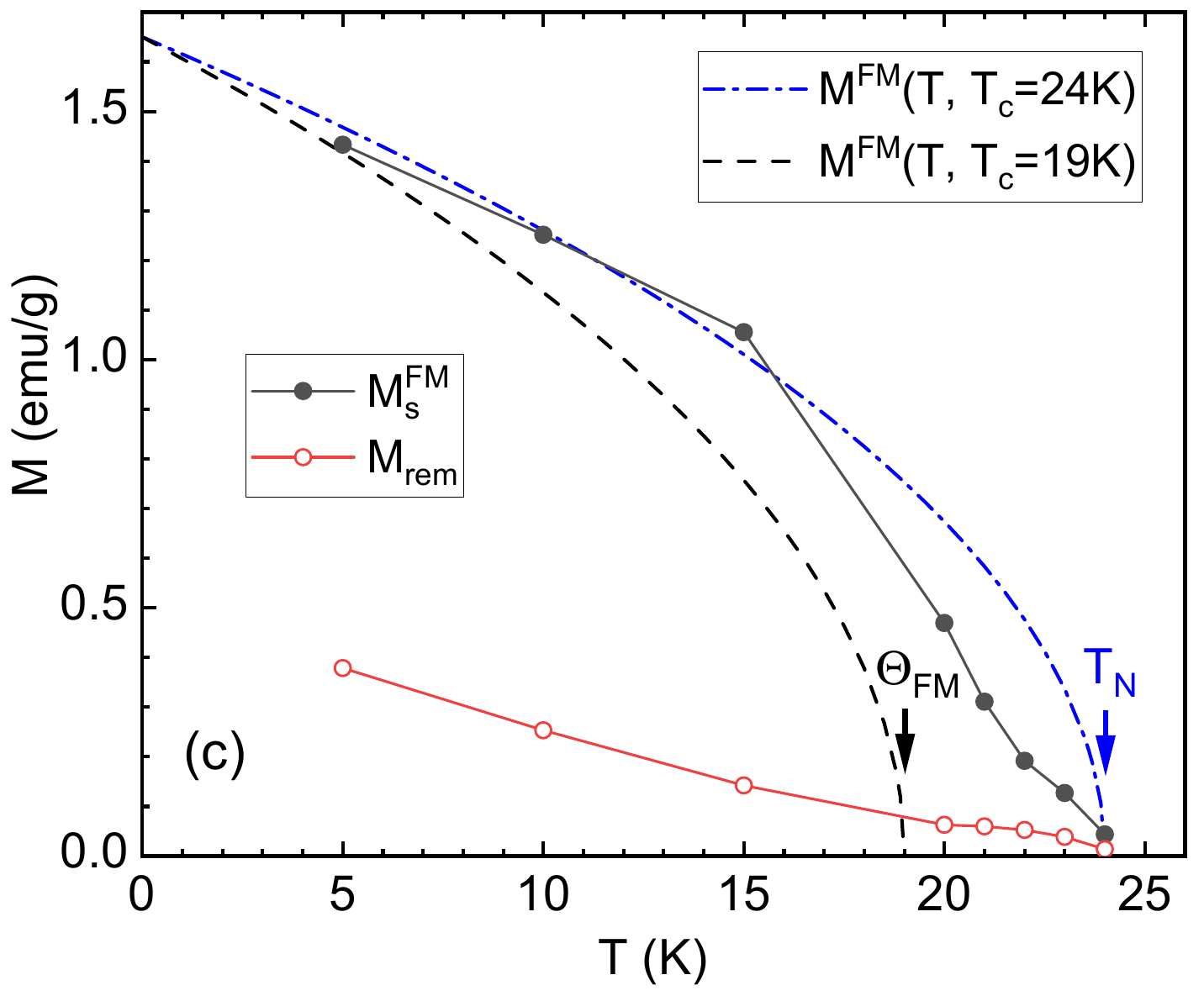}
		\caption{(a) DC-magnetization $M(H)$ for EuSn$_2$As$_2$ crystal at $T=2$K 	for two field orientations. Saturation magnetization $M_s\approx 60 {\rm \,emu/g}$ (i.e. $\approx 7\mu_B$/f.u.). Arrow show saturation field $H_s\approx 3.6$T for $H\|(ab)$.  The raw data are not corrected for demagnetizing factor $N_i$. For our typical sample $2\times 1\times 0.2$mm, $N_{ab}\approx 0.16$ for  $H\|(ab)$ and $N_c\approx 0.69$ for $H\|c$. Correspondingly, the magnetic field scale  should be corrected by $\sim 6\%$ for $H\|(ab)$ and by $\sim 19\%$ for $H\|c$.
	(b) Magnified low-field interval of the  $M(H)$ curves 
	 for the 2mg-piece cutout from the same bulk crystal measured at eight  temperatures 
	 for $H\|ab$  orientation. The inset:   $M(H)$ dependence at $T=5K$ after subtracting the AFM magnetization slope. The inset also shows graphical definitions of the extrapolated to $H=0$ saturation magnetization  $M_s^{\rm FM}$    and remanence magnetization $M_{\rm rem}$. 	 
	 	 (c)	 Temperature dependences of  $M_s^{\rm FM}$ and $M_{\rm rem}$.
	 Vertical arrows point at $\Theta_{\rm FM}$ and $T_{\rm N}$. 
	{\cb Dashed and dash-dotted curves show the model  $M_{def}(T)=M_s^{\rm FM}$  dependence, Eq.~(\ref{eq:landau}), for two critical temperature values.}
 		}
		\label{fig:M(H)}
	\end{figure*}

DC magnetization curves measured to higher fields are shown in  Fig.~\ref{fig:M(H)}a; 
at first sight these curves are typical for the easy-plane AFM crystal \cite{macNeill_PRL_2021}. They are  fully consistent with our previous measurements on similar samples \cite{golov_JMMM_2022, pervakov_PRB_2026};  such data  were also  reported  in  Refs.~\cite{arguilla_InChemFront_2017, pakhira_PRB_2021, li_PRX_2019, chen_ChPhysLet_2020}.
Magnetization saturates in fields above $H_s$  due to  the  full spin polarization: $H_s\approx 4.7$\,T, when $H\|c$,   and  $H_s\approx 3.4$T   when $H$ lies in the easy magnetization $ab$-plane {\cb (more detailed $M(H)$ curves for various temperatures are presented in Fig.~9 of SM \cite{SM})}.  

For $H\|c$ the magnetization curve is almost linear, however when  field is applied in the $ab$ plane,  $M(H)$ exhibits  a sample-dependent weak  nonlinearity in low fields less than $\approx 0.03\,$T. 
In  Fig.~\ref{fig:M(H)}a we show again the 
non-linear $M(H\|ab)$ dependence taken with the same sample, as  in  Fig.~\ref{fig:chi(T)}a. 
Similar nonlinearity  of $M(H\|ab)$  in low fields may be noticed also in earlier published data of Refs.~ \cite{pakhira_PRB_2021, li_PRB_2021, chen_ChPhysLet_2020, li_PRX_2019, pervakov_PRB_2026, golov_JMMM_2022}. 

In order to explore the unforseen $M(H)$ nonlinearity for $H\|(ab)$, we have measured magnetization in more detail. 
Figure~\ref{fig:M(H)}b presents the magnified ($\times 100$)  low-field range of the  DC magnetization $M(H)$ curves. In low fields, from $-0.03$ to $+0.03$\,T, the in-plane magnetization $M(H_{ab})$ is non-linear and displays a  hysteresis of a 
ferromagnetic type. 
The $M(H)$ curve in Fig.~\ref{fig:M(H)}b is a superposition of the typical ferromagnetic magnetization curve on top of the host lattice AFM linear magnetization (whose slope is temperature independent). Obviously, namely this hysteresis is the origin of the low-field nonlinearity  of $M(H)$ that may be seen in Fig.~\ref{fig:M(H)}a for $H\|ab$. 
 The magnetization hysteresis disappears at $T \approx 24 - 25$\,K according to the temperature dependences of  the width and height of the hysteresis loops.
  Figure~\ref{fig:M(H)}c  shows  the saturation magnetization $M_s^{\rm FM} (T)$ and remanent magnetization $M_{\rm rem}(T)$;  the coercitivity decays similarly to  the latter ones.

  {\cb Temperature dependence of the saturation ferromagnetic magnetization $M_s^{\rm FM}$ is shown in Fig.~\ref{fig:M(H)}c.
  	For comparison we also plotted the textbook $M(T)$ dependence for ferromagnets   in the vicinity of a critical temperature 
  	$T_c = 19$  or $24$\,K \cite{landau}:
  	\beq
  	M^{\rm FM}(T)= M_0 \sqrt{ (T_c-T)/T_c}
  	\label{eq:landau}
  	\eeq
  }
    In Fig.~\ref{fig:M(H)}c, the fitting parameter    $M_0=0.165$\,emu/g  (i.e., $M_{\rm s}^{\rm FM} (T)$ value extrapolated to $T=0$ )	is 37 times smaller than the AFM saturation magnetization $M_s$ (Fig.~\ref{fig:M(H)}a).  We discuss this result in subsection \ref{sec:discussion-B}.
	The attentive reader may notice in Fig.~\ref{fig:M(H)}a  a slight nonlinearity in the magnetization also along the $c$-axis 
	in the interval of $H=0.5 -2$T. It is typical for the easy plane AFM materials (see, e.g. Fig.~10 of Ref.~\cite{pakhira_PRB_2021}, Fig.~3c of \cite{li_PRB_2021}, Fig.~2b of \cite{chen_ChPhysLet_2020}, Fig.~3 of \cite{arguilla_InChemFront_2017}) and reflects spin reorientation from the in-plane to the $c$-axis (spin canting  and spin-flop) 
	as field increases. The spin reorientation process extends up to 15-60 times higher fields than $H_s^{\rm FM} \sim 30$mT and therefore is irrelevant to the studied ferromagnetic-like features. This issue is beyond the scope of our paper and is considered in several preceding works. 
	Another kind of nonlinearity in the $c$-axes magnetization may be seen in Fig.~11d of Ref.~\cite{pakhira_PRB_2021} in low
	fields $(10 – 20)$mT; it manifested there as a deviation of the  $M/H$  field dependence from the horizontal line. 
	This feature, however, persisted up to 100\,K, well above $T_N$; therefore it has an origin, different from the studied here  ferromagnetism emerging from planar defects.

 \section{Transmission electron microscopy investigations }
\label{sec:TEM}
\begin{figure*}
\includegraphics[width=0.4\linewidth]{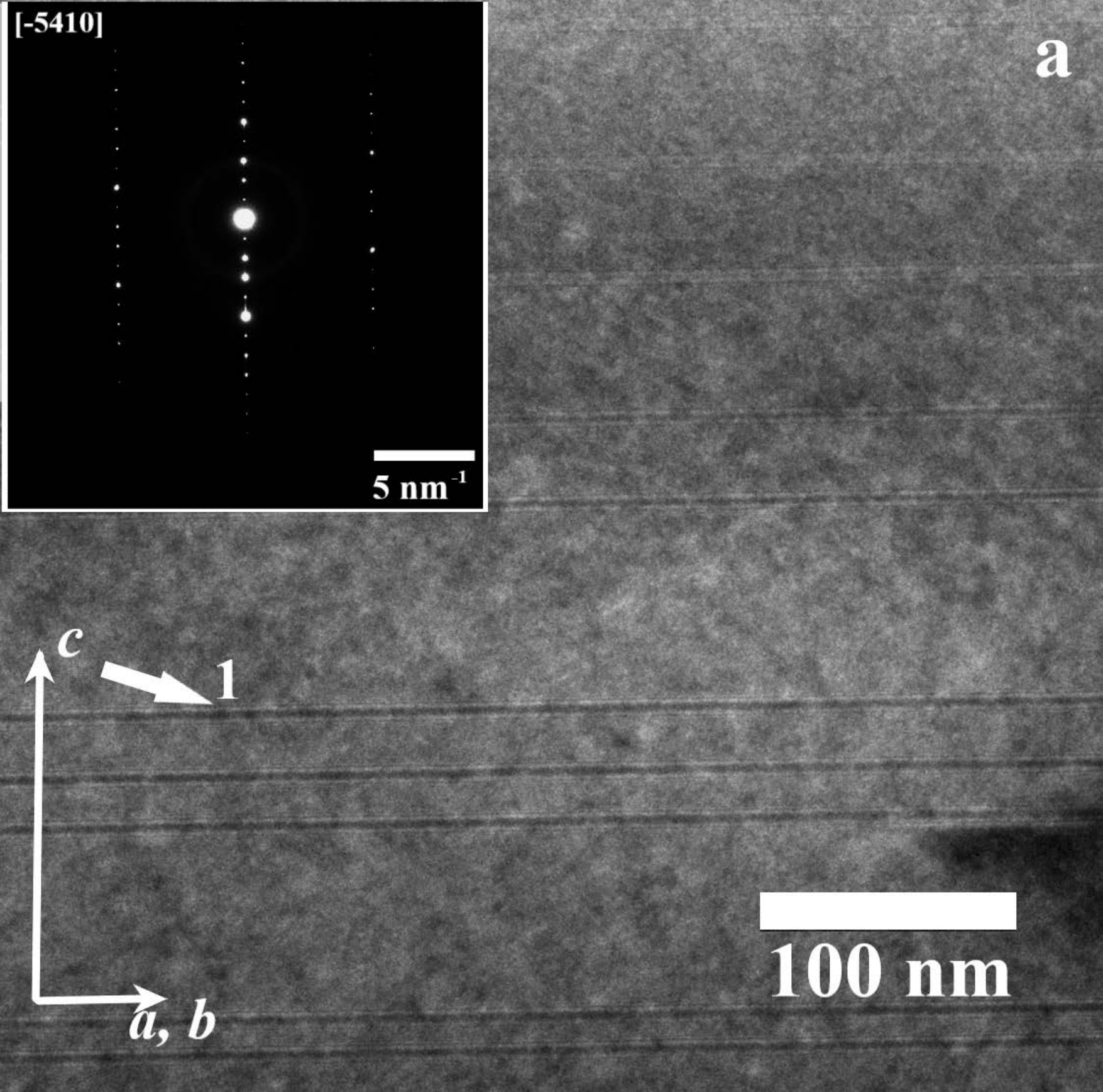}
\includegraphics[width=0.47\linewidth]{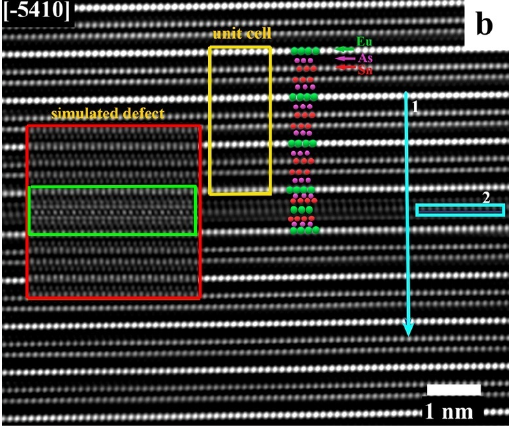}\\
\includegraphics[width=0.85\linewidth]{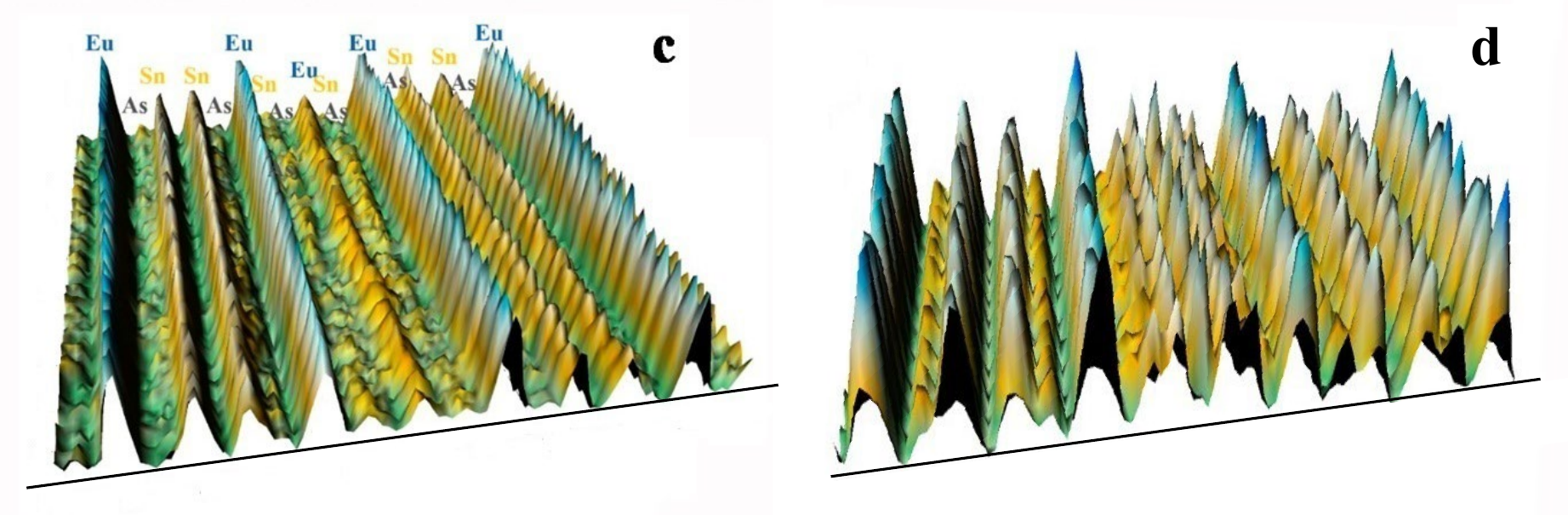}
\includegraphics[width=0.85\linewidth]{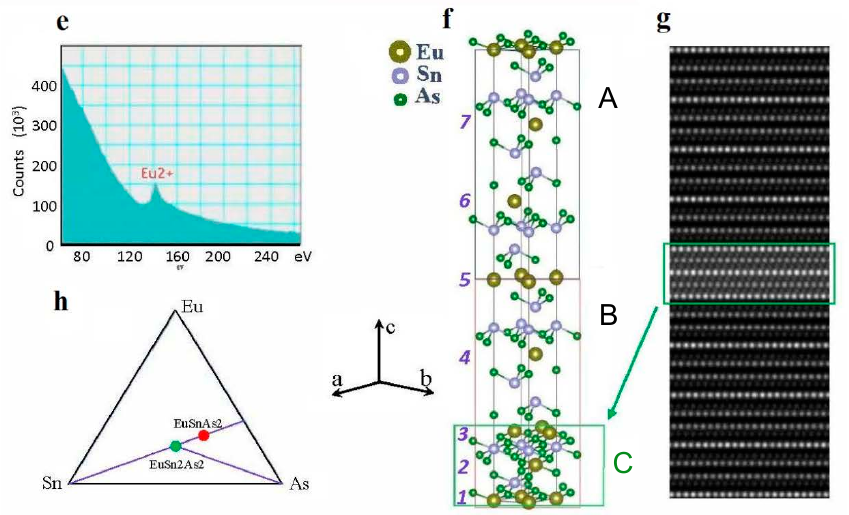}
\caption{(a) BF TEM image of  EuSn$_2$As$_2$ single crystal with planar defects  
(one is marked with the white arrow \#1). Inset: SAED obtained in $[\bar{5}410]$ zone axis.
(b) FFT filtered HR HAADF STEM image of the crystal lattice. Green balls designate Eu atoms, red balls - Sn, and violet - As. The EuSn$_2$As$_2$ unit cells is marked with  yellow frame,
blue arrow ``1''  depicts the  EDXS analysis line and blue rectangle ``2'' - of the EELS analysis. 
{\cb     The simulated image in the red frame of panel  (b) models the initial experimental image prior to FFT analysis;}
(c) the intensity maps obtained from HR HAADF STEM experimental image of defect area and (d) intensity map - from simulated image; (e)  EELS spectra obtained from the area marked with blue rectangle 2 on panel (b); 
 (f) DFT simulated crystal structure model Eu$_7$Sn$_{12}$As$_{14}$  of the defect supercell,  including the transition layers 
 and defect monolayer itself; 
(g) {\cb
 a magnified fragment of the red framed   area of HR HAADF-STEM image   overlaid with a simulated image of the crystal structure containing a defect;}
(h) a simplified triangle phase diagram (convex hull) of the Eu-Sn-As system, showing the two major phases (``122'' and ``112'').
}
	\label{fig:TEM}
\end{figure*}

 In total, about ten  lamellas, $15\times 15\mu$m  wide in $(ac)$-plane, and (10-15)\,nm thick, have been prepared and investigated using TEM; in all of them we found planar defects whose volume concentration varied within  $2 - 4\%$, as estimated from the width of the defect stripes (see Fig.~\ref{fig:TEM}a and also SM \cite{SM}).
In one of the lamellas, investigated in most detail,   the share of defect volume is about 3\%.
The bright field (BF) TEM image of  EuSn$_2$As$_2$ single crystal (see Fig.~\ref{fig:TEM}a)  revealed the presence of planar defects (marked with arrow \#1) which are the focus of our study. 
Selected area electron diffraction (SAED) pattern (upper left corner inset of Fig.~\ref{fig:TEM}a) 
indicates that the specimen was observed along the   $[\bar{5}410]$ 
zone axes. Here and below the structure is indexed with respect to hexagonal axes  (see Fig.~\ref{fig:1}b). High resolution (HR) HAADF STEM image (Fig.~\ref{fig:TEM}b) and SAED pattern with streaks parallel to $c^*$-axis  unambiguously demonstrate that the habit plane of the defects is parallel to the \{0001\} crystal planes of EuSn$_2$As$_2$.  
	
The pristine  EuSn$_2$As$_2$ unit cell (highlighted in Fig.~\ref{fig:TEM}b with a yellow frame) exhibits three blocks, in accordance with the lattice structure shown in Fig.~\ref{fig:1}a. Each block 
in HR HAADF STEM images demonstrates the 
chains of spots (Fig.~\ref{fig:TEM}b): (i) the brightest chain corresponds to 
columns of Eu atoms, (ii) less intense are the double chains of columns of Sn atoms,
and (iii) chains of columns of As atoms are nearly visible in the images but appear in the 
intensity map (Fig.~\ref{fig:TEM}c).
From here on, the atomic chains are the rows of atoms perpendicular to the electron beam, and atomic columns are those parallel to it. It has recently been established that the intensity of the atomic column image is proportional to the atomic number  $Z^\gamma$ ($\gamma=1.6 - 2$) and depends nearly linear on the number of atoms in the column \cite{vandenBroek(2012), vanaert(2011)}.
	
\begin{figure*}
\includegraphics[width=360pt]{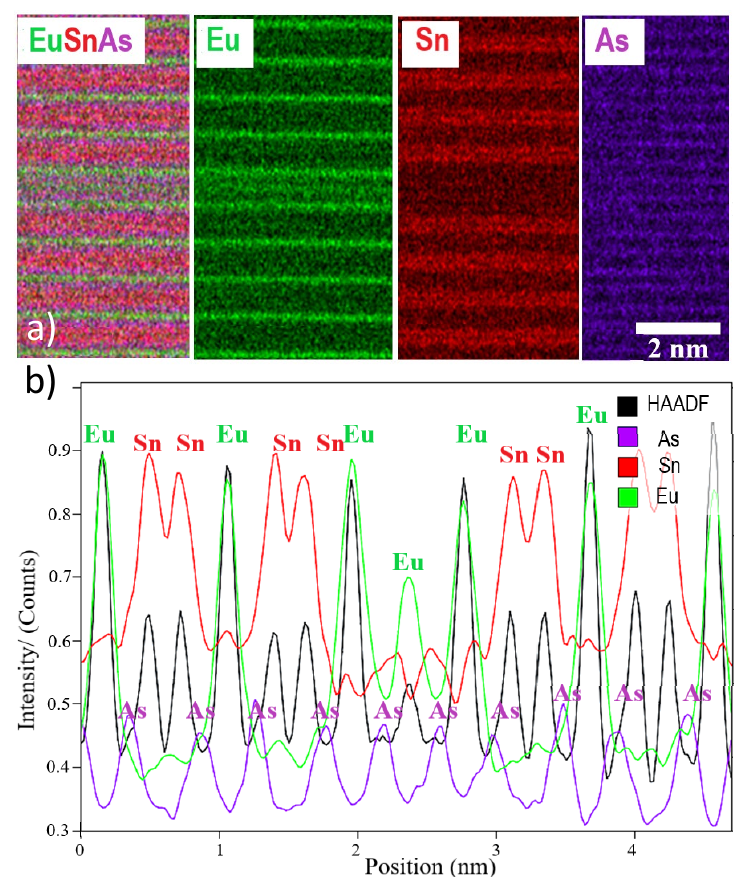}
\caption{ (a) EDXS Chemi STEM elemental maps in a region including planar defect;
(b) EDXS Chemi STEM elemental line profile along blue arrow 1 in Fig.~\ref{fig:TEM}b.
}
\label{fig:EDX map}
\end{figure*}

In addition to the regular structure of the atomic layers (e.g., framed in yellow in Fig.~\ref{fig:TEM}b), one can see the defect layers.
  {\cb The simulated image in the red  box of Fig.~\ref{fig:TEM}b models the initial experimental image prior to FFT analysis; it includes both, several regular layers and the defect layer.  An example of the  modeled image of the defect layer is framed in the green box (see Fig.~\ref{fig:TEM}b and also  Fig.~7 in \cite{SM}).} The contrast in the defect area 
differs strongly: there are three chains of brighter spots 
with slightly higher intensity in the central row  that is also evident from 
the intensity maps (Fig.~\ref{fig:TEM}c). 
The EELS analysis of the central chain demonstrates the presence of Eu (Fig.~\ref{fig:TEM}c). 
This is confirmed by the results of the EDXS: the presence of Eu atoms chain is unambiguously revealed by the elemental mapping (Fig.~\ref{fig:EDX map}a) together with line scan across the crystal layers (Fig.~\ref{fig:EDX map}b). 

Analyzing the data for 10 samples, we found that 
 the distance between individual planar defects  along the $c$-axes
ranges   from  6 to $70\,$nm, whereas the  length of the defect along $a$ (or $b$) axes,  exceeds 600\,nm (the estimate is limited by the TEM field of view, see Fig.~7 in SM \cite{SM}).  In BF TEM images one can also see dislocations in the planar defects termination areas (Fig.~7 of \cite{SM}).

The EDXS data (Fig.~\ref{fig:EDX map}b) demonstrates that
the intensity of Eu peak in the middle of the planar defect is significantly lower  than that of two Eu peaks obtained from the defect-free EuSn$_2$As$_2$ areas. The Sn peaks in the defect area are noticeably lower, than those in the defect-free areas. Close inspection of the images 
 indicates that the crystal lattice parameter of the planar 
 defect along the $c$ - axis is $c_{\rm def}= 0.8$\,nm, 
in contrast with  the Eu-Eu distance $c/2 \approx 1.32$\,nm for the defect-free lattice \cite{pakhira_PRB_2021, pervakov_PRB_2026}. 
The part of EELS spectrum for the Eu N$_{4.5}$-edge demonstrates that the oxidation state of Eu has conventional value of $2^+$ \cite{mundy_APL_2014, kim_PRB_2021}. Thus, if we consider the fact that half of the Sn-atoms are deficient in the defect, the local compound in the defect area takes the following chemical form: Eu$^{2+}$Sn$^{4+}$As$^{3-}_2$ (cf. the chemical formula for the regular structure is Eu$^{2+}$Sn$^{2+}_2$As$^{3-}_2$). 
 
 The ternary phase diagram of the Eu-Sn-As system (Fig.~\ref{fig:TEM}h) in a 
relevant temperature range was carefully studied and, using the stable crystalline phase detection scheme \cite{wang_NPJCM_2021}, EuSnAs$_2$ was identified as the only stable compound that meets the chemical criteria. Based on the observed geometric parameters of the defect, the ternary phase diagram,  and taking into account our DFT calculations (see section \ref{sec:DFT}), we propose the crystal phase EuSnAs$_2$ (Fig.~\ref{fig:TEM}f) as a single layer defect composition. Noteworthy, the EuSnAs$_2$ phase is adjacent to EuSn$_2$As$_2$ on the ternary phase diagram (Fig.~\ref{fig:TEM}h). 

The planar defect lattice structure  was constructed by including single layer of the EuSnAs$_2$ compound having face-centered cubic symmetry (space group $Fm\bar{3}m$) \cite{QM-Database}. To match the original trigonal structure (Space group 
$R\bar{3}m$, a primitive unit cell of $Fm\bar{3}m$ structure with the basis $a=(0, 0.5, 0.5)$, $b = (0.5, 0, 0.5)$, $c = (0.5, 0.5, 0)$ was rotated and transformed by the matrix $\left((2,-1,-1), (-1, 2, -1), (2, 2, 2)\right)$. As we found, the 7 layers of the resulting defect structure match the EuSn$_2$As$_2$ cell.  The final supercell with 
the planar defect has the exact formula Eu$_7$Sn$_{12}$As$_{14}$ with 7 Eu atoms per unit cell, 
whereas the pristine 3-layer host lattice structure with AFM order may be considered as Eu$_6$Sn$_{12}$As$_{12}$.

The crystal structure model is presented in Figs.~\ref{fig:TEM}f,g,  where  the top part (block {\bf A}) represents the unit cell of EuSn$_2$As$_2$, containing 3 Eu atoms, 6 As atoms and 6 Sn atoms. (taking into account the neighbor unit cells). The 
second elementary cell {\bf B} is modified and 
borders the defect layer  {\bf C}  (the latter is framed by a green rectangle) embedded between the two Eu layers. It doesn't matter between which layers it is embedded;  on the figure it is placed at the bottom for ease of working with the cell and numbering of atoms {\cb (the counting of europium atoms is presented in  SM \cite{SM}.)
} 

In order to verify our construction, we attempted to simulate  the HR HAADF STEM image of the supercell containing planar defect with a number   of different specimen thickness. 
However, these simulations do not demonstrate perfect agreement with experimental images 
 in the relevant range of thickness between 10\,nm and 100\,nm. Specifically, two features in the simulations do not match the data: (1) The chain of Eu atoms in the defect area exhibits too high intensity in the simulated images. (2) The contrast  of Sn and As chains is also different from that in the defect-free region. In order to  overcome this discrepancy we  assumed that in most cases the defect structure - EuSnAs$_2$ overlaps with EuSn$_2$As$_2$ crystal forming a ``two-layered''  heterostructure in the direction parallel to the electron beam  (see Fig.~\ref{fig:spatial arrangement}).

Here, for brevity, we present only the details of the simulated structures that gave acceptable agreement 
with the experimental HR HAADF STEM image.   The best fit was obtained for the thickness (along the beam direction) of the EuSnAs$_2$ defect layer of  5\,nm and the EuSn$_2$As$_2$ layer of 13\,nm. 
{\cb Since directly resolving the As planes in the image   is challenging, an additional experimental intensity profile was obtained, as shown  in Fig.~\ref{fig:EDX map}b. In this profile, peaks corresponding to the 7 atomic layers of As can be clearly seen.}

 The simulated image   is outlined in red in Fig.~\ref{fig:TEM}b
and the intensity map obtained from this image is shown in Fig.~\ref{fig:TEM}d.
Except for the acceptable peak intensities of Eu and Sn, additional weak peaks 
from the overlapping As and Sn chains of the defect and the ideal crystal coincide 
with the experimental image (compare Fig.~\ref{fig:TEM}c and Fig.~\ref{fig:TEM}d).

\section{DFT calculations}
\label{sec:DFT}
\subsection{Defect band structure and magnetic properties}

\begin{figure*}[ht]
	\centering
	\includegraphics[width=0.45\textwidth]{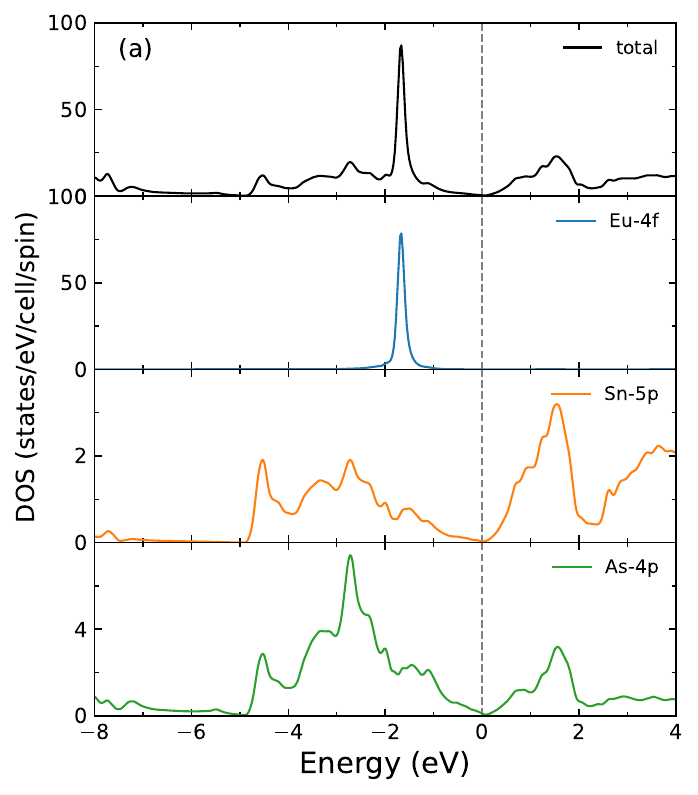}
	\includegraphics[width=0.45\textwidth]{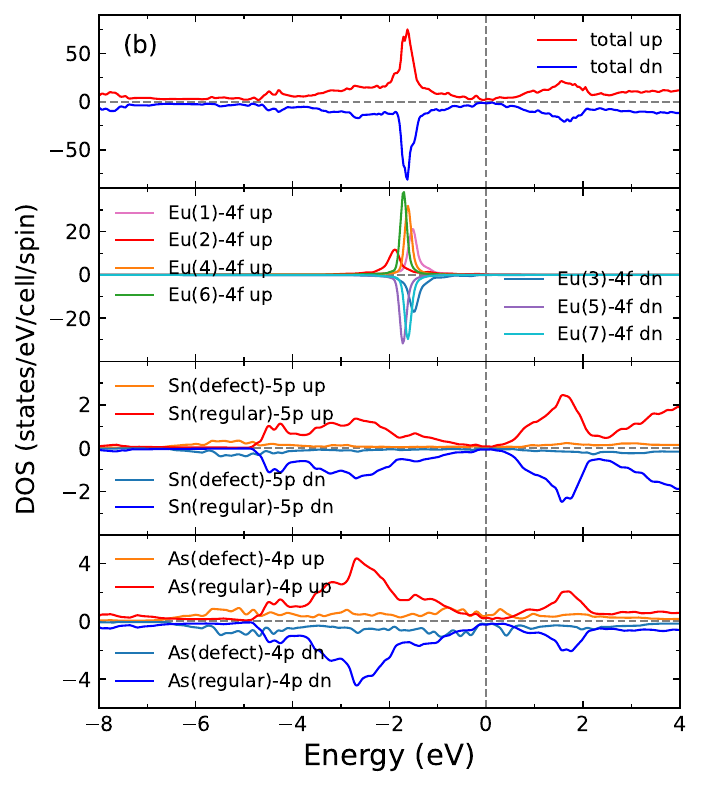}
	\caption{GGA+U total and partial DoS for EuSn$_2$As$_2$ (a), and for Eu$_7$Sn$_{12}$As$_{14}$ (b). For EuSn$_2$As$_2$, the density of states with spin up and spin down coincide.
 In case of Eu$_7$Sn$_{12}$As$_{14}$, DoS also coincides for atoms Sn(regular) and As(regular) located away from the planar defect. The Fermi level corresponds to the zero energy. 
	} 
	\label{fig:7}
\end{figure*}

In order to understand the nature and characteristics of the EuSnAs$_2$ planar defect and its interaction with the regular crystal lattice, we performed DFT+U calculations of Eu$_7$Sn$_{12}$As$_{14}$ (see computational detail 
in Section 2). As we have shown above (Fig.~\ref{fig:TEM}f), the supercell of the planar defect includes 
an odd number (7) of Eu atoms. Therefore, the presence of a planar defect in the crystal structure leads to the 
appearance of an uncompensated magnetic moment per unit cell. 

As a result of this decompensation, the Eu-4f density of states (DOS) of Eu atoms from regular crystal lattice (atoms 1, 3 - 7 enumerated in Fig.~\ref{fig:TEM}f) have maximum at slightly different energy $-1.5, -1.6, -1.7$~eV (Fig.~\ref{fig:7}b). 
It is similar for up and down spin component. For the  Eu atoms 
in the center of the {\cb defect supercell }
 (part B on Fig.~\ref{fig:TEM}f) the peak of DOS is located at $-1.9$~eV (Fig.~\ref{fig:7}b). For the regular crystal structure EuSn$_2$As$_2$ with AFM configuration, the total DOSes for up and down spin component coincide (Fig.~\ref{fig:7}a). The peak of Eu-4f DOSes for different atoms is located at the same energy $-1.7$~eV. The Sn and As atoms in the {\cb defect supercell} (atoms inside part B on Fig.~\ref{fig:TEM}f) have a small magnetic moment. Also, the contribution to DOS from  these Sn  and As atoms in the {\cb defect 
  supercell } shifts down in energy to $-6$\,eV (Fig.~\ref{fig:7}b). In contrast, DOS of Sn (regular) and As (regular) atoms from the regular crystal lattice have the lower boundary at $-5$\,eV. 

\begin{figure}[ht]
	\includegraphics[width=220pt]{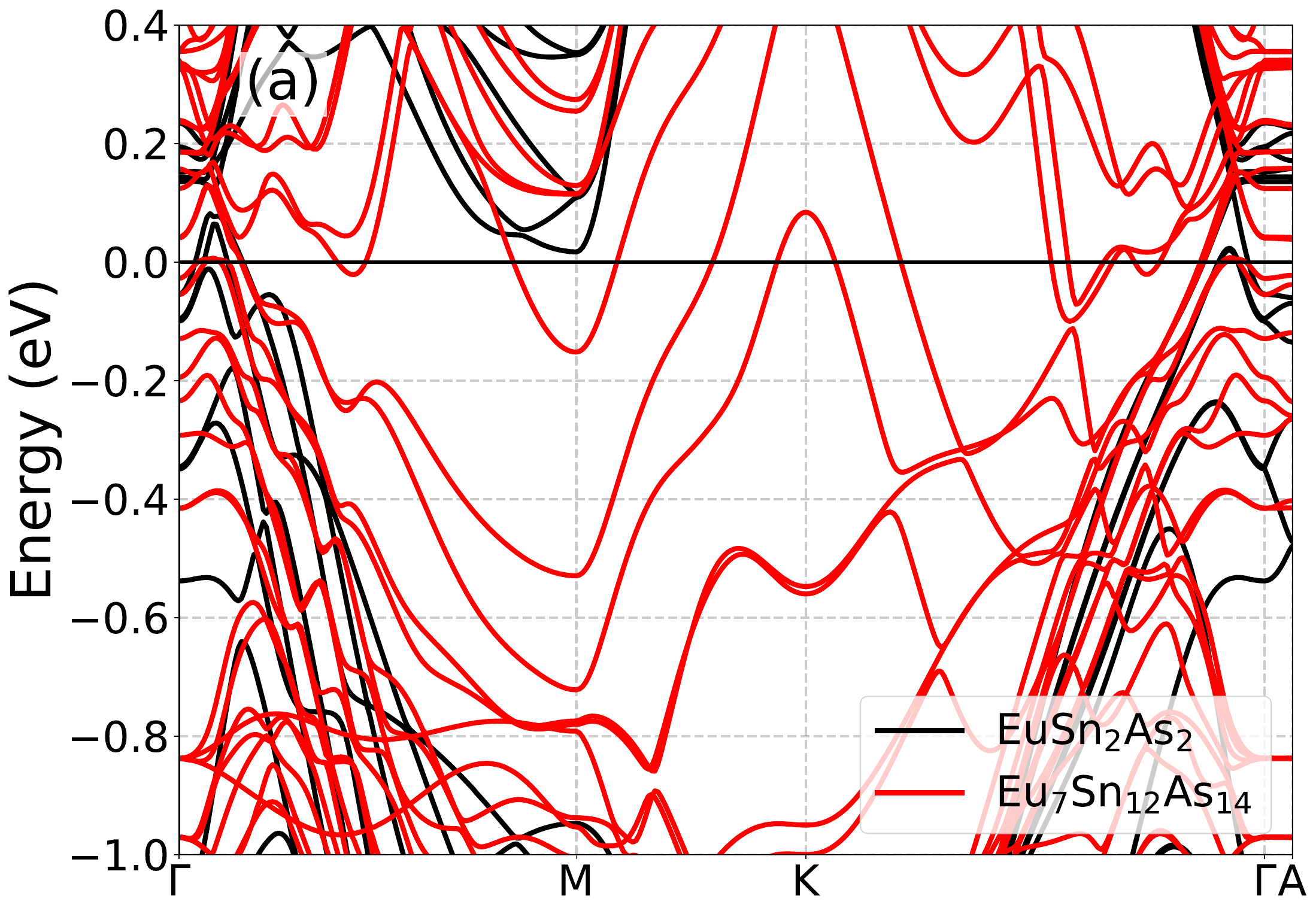}
	\includegraphics[width=220pt]{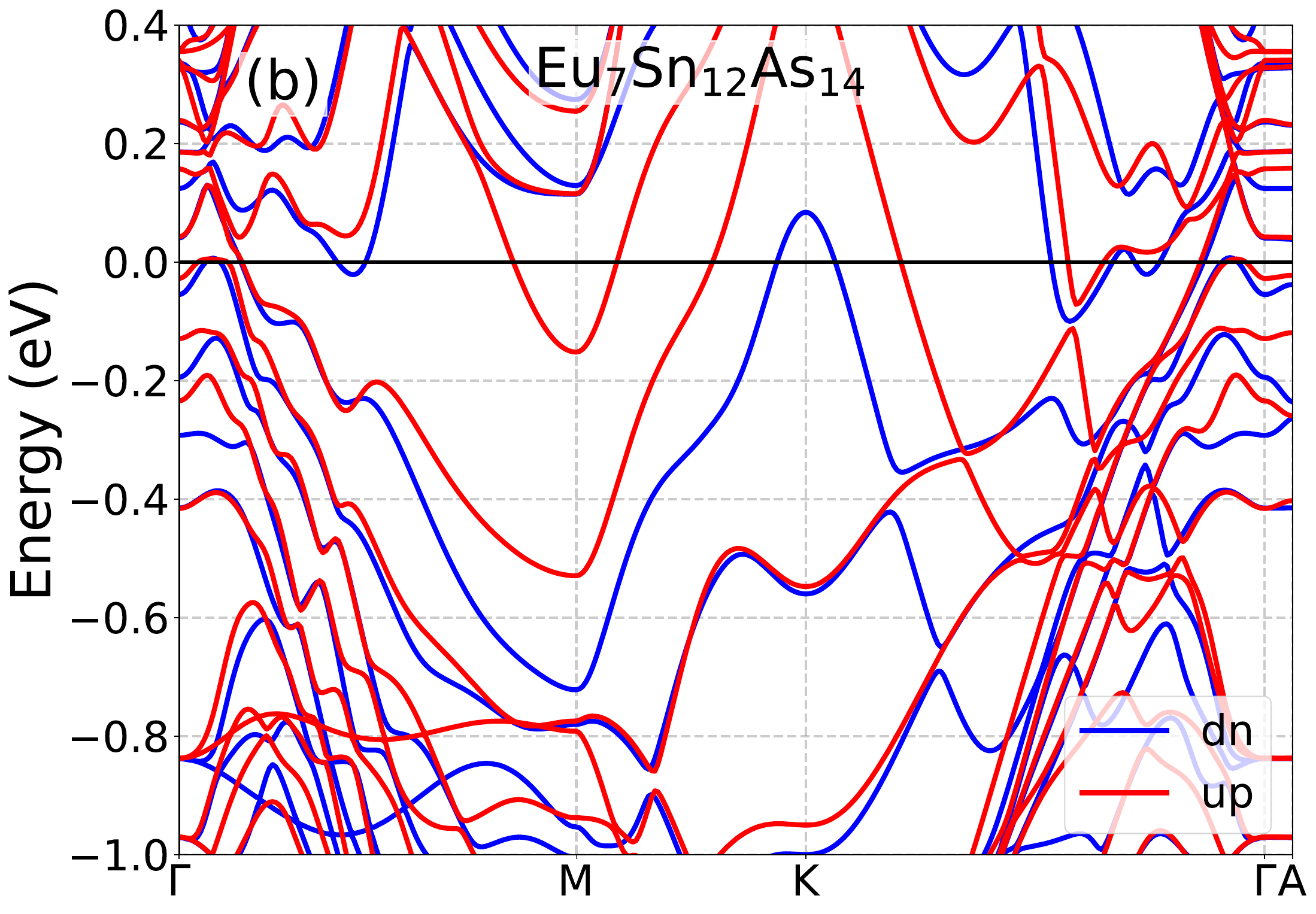}
	\includegraphics[width=220pt]{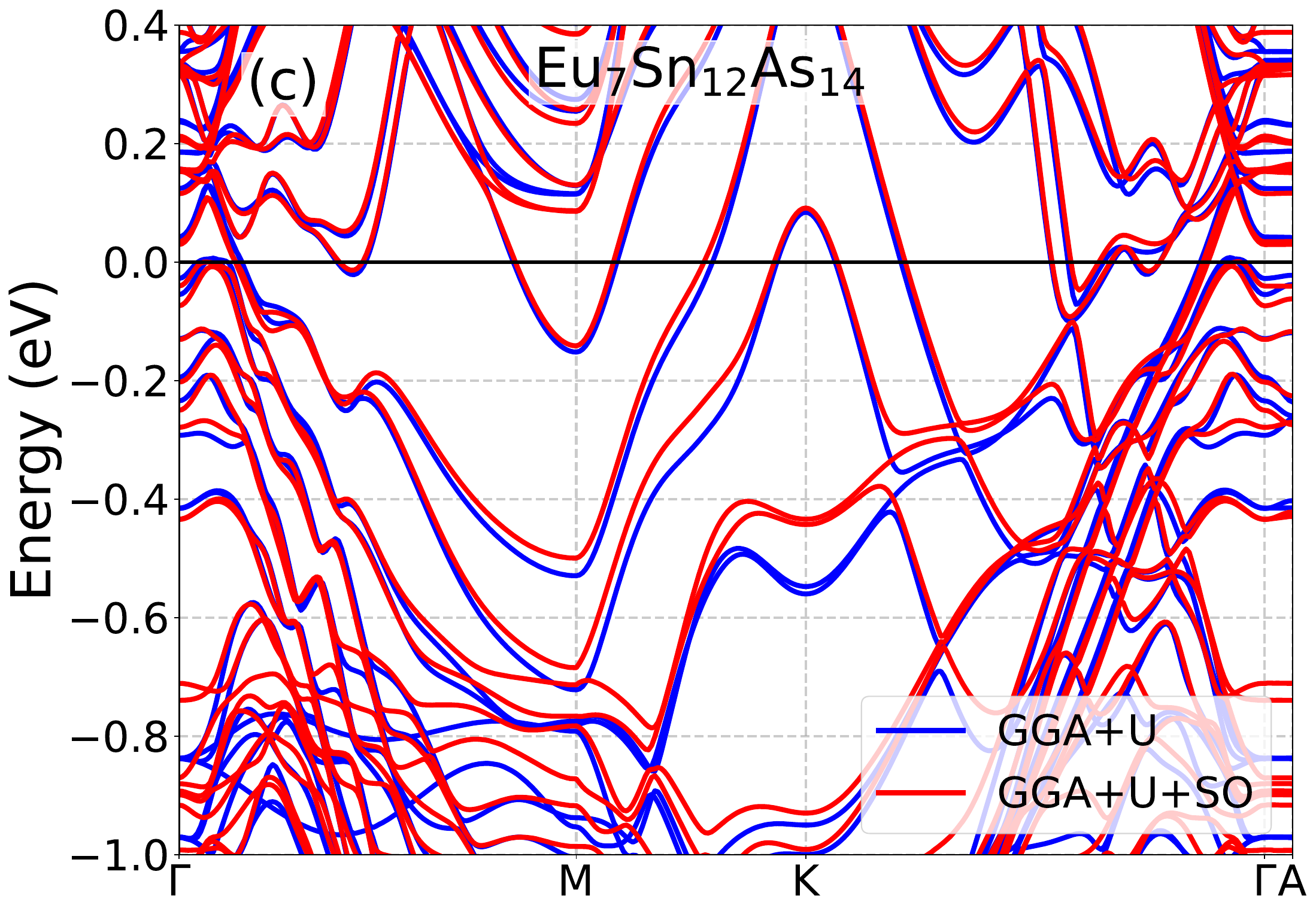}	
	\caption{Band structure (a) for the  lattice with the  planar defect of Eu$_7$Sn$_{12}$As$_{14}$ (red lines) compared to the bulk defect-free structure of EuSn$_2$As$_2$ (black lines) obtained in GGA+U. 
		(b) GGA+U band structure for Eu$_7$Sn$_{12}$As$_{14}$ resolved by spin. 
		(c)The effect of spin-orbit interaction accounting to the band structure of Eu$_7$Sn$_{12}$As$_{14}$. 
	} 
	\label{fig:8}	
\end{figure}

We now show how the presence of a defect in the crystal structure affects the band structure. Comparison of the band structure of the {\cb supercell }  Eu$_7$Sn$_{12}$As$_{14}$ (red lines) with that of the bulk EuSn$_2$As$_2$ (black lines) is presented in Fig.~\ref{fig:8}a. 
The Eu$_7$Sn$_{12}$As$_{14}$ bands cross the Fermi level in the vicinity of the M point and along M-K, K-$\Gamma$ directions of the Brillouin zone. In case of  ESA,  the bands cross the Fermi level only near the $\Gamma$ point. These new bands in Eu$_7$Sn$_{12}$As$_{14}$  are mainly associated with the As-4p states of atoms, and are located in the planar defect region (part C in Fig.~\ref{fig:TEM}f). The Eu$_7$Sn$_{12}$As$_{14}$ extra bands should make little or no contribution to the ARPES measurement results  due to the small volume fraction of the defects  ($\sim 3\%$) in the investigated single crystals. However,  upon cooling the uncompensated Eu-magnetic moments are expected to experience  ferromagnetic-type  ordering, which may be revealed, e.g. in magnetization and magnetic resonance measurements.

Figure~\ref{fig:8}b shows the Eu$_7$Sn$_{12}$As$_{14}$ bands for each spin component. 
Here, the bands with spin-orbit interaction (SO) included are depicted by red lines,  and 
without SO - by blue lines.  The splitting of bands with opposite spin direction near the Fermi level 
with spin-orbit interaction included are depicted
by red lines, and without SO - by blue lines. One can see that spin-orbit interaction only slightly shifts the bands and almost doesn't split the bands for Eu$_7$Sn$_{12}$As$_{14}$ (see Fig.~\ref{fig:8}c). 

As a result of the DFT+U calculations, we obtained the magnetic moments on each of 7 Eu atoms (in Bohr magnetons) 
in the unit cell of the model crystal structure Eu$_7$Sn$_{12}$As$_{14}$ as follows: 6.73, 6.70, -6.73, 6.81, -6.81, 6.81, -6.81 (the sequence is indicated  from bottom to top along the $z$-axis in Fig.~\ref{fig:TEM}f).  Total magnetic moment DFT+U 
 is $6.70\mu_B$ per formula unit of defective monolayer, and $6.77\mu_B$ per f.u. of the supercell  {\cb (for more detail, see Sec. VIII of SM \cite{SM}).} To estimate the total magnetic moment of a real bulk macroscopic sample, the DFT+U calculated magnetic moment is multiplied by volume fraction of the monolayer defects in the bulk, $\approx 3\%$  {\cb (see Sec. V of SM \cite{SM})}. Thus, the average 
 magnetic moment of the bulk sample per unit volume is about $0.20 \mu_B$.

\subsection{Brief summary of the computation results}
\label{sec:discussion-B}
\subsubsection{Defect formula and unit cell}
	From the 	high resolution TEM investigations we revealed the presence of planar defects in a high quality EuSn$_2$As$_2$ crystals. From the DFT calculations we have identified the 
	structure and composition of the planar defect. {\cb The defect monolayer} is found to incorporate an extra and misplaced row of Eu atoms, 	and has the local composition  EuSnAs$_2$ due to missing half of Sn atoms. This compound with cubic crystal structure is stacking  by its (111) plane with  trigonal structure of the parent EuSn$_2$As$_2$.  The extended elementary cell (supercell) of the defective transition layer contains 33 atoms and has a total formula Eu$_7$Sn$_{12}$As$_{14}$. 
  
\subsubsection{Magnetization}
{\cb\underline {Uncompensated local magnetic moments. }\\
	Since both the defect monolayer EuSnAs$_2$ and the supercell  Eu$_7$Sn$_{12}$As$_{14}$ contain odd numbers of Eu atoms,  the defect 
can be  	expected to have a non-zero magnetic moment. 
Indeed, within DFT+U  we found that   the  uncompensated magnetic moment 
is  $\approx 6.7\mu_B$  per unit cell of  the defect monolayer EuSnAs$_2$. 
Provided the uncompensated moments are ferromagnetically ordered in the bulk sample, 
this value together with the defect volume fraction would 
result  in the sample magnetization $\approx 0.2 \mu_B$/unit volume)}. This estimate  is
consistent with our magnetization measurements, thus indirectly confirming the ferromagnetic-type ordering of uncompensated Eu moments.
 Indeed, from Fig.~\ref{fig:M(H)}b  one can see that the ratio  $M_s^{\rm FM}/M_s\approx 0.027$. Given $M_s \approx 7 \mu_B$/f.u.  \cite{pakhira_PRB_2021, arguilla_InChemFront_2017, li_PRB_2021, chen_ChPhysLet_2020}, we obtain $M_s^{\rm FM} \approx 0.19\mu_B$/f.u., close to the  above DFT theoretical estimate $0.2\mu_B$ (within the  uncertainty of the volume fraction of planar defects ($3 \pm 1\%) $.

{\cb
\underline{Mechanisms of magnetic ordering in the host}  \underline{EuSn$_2$As$_2$ crystal and in the planar defects}.\\
	DFT calculations of Ref.~\cite{sun_SciChina_2021} provide an estimate of ferromagnetic type in-plane exchange coupling $J_{\rm FM}=-1.9$\,meV between the nearest-neighbor Eu atoms in the ESA 	lattice, separated by a distance of 4.2\AA,  in the $(ab)$ plane. 
	This IP coupling provides the short-range  in-plane FM ordering of Eu moments starting from high temperatures.
	
Evaluating various mechanisms behind the  emergence of magnetic ordering in the defect, we find that the major contribution also comes from exchange interaction.  Indeed, the in-plane distance between the nearest neighbor Eu atoms, 4.2\AA  	\cite{sun_SciChina_2021}  is about the same for $(ab)$-plane of the regular trigonal  ESA  
lattice and for the diagonal (111) plane of the cubic EuSnAs$_2$ lattice of the planar defect.  

 In contrast,  the interplane exchange coupling in the regular ESA  lattice has the antiferromagnetic sign $J\approx +0.2$\,meV.  This is the result of  a subtle interplay between various large contributions to interaction with different signs,  such as  direct exchange (FM-type), RKKY (FM type), and superexchange (AFM type). 

For the defect compound EuSnAs$_2$, the enhanced valence of tin, missing half of Sn-atoms and the smaller ionic radius  of the Sn(+4) ions separating the Eu-layers in the defect monolayer lead to the smaller Eu-Eu interplane distance  (about 0.8nm instead of 1.3nm in the regular ESA lattice). As a result, the interplanar exchange interaction of Eu atoms in the defective  monolayer also acquires a ferromagnetic sign and leads to the orientation of the uncompensated magnetic moment  parallel to the magnetic moment of the adjacent layer of Eu atoms in the regular ESA lattice.
The above reasoning provides a qualitative explanation of the result of our DFT calculations.} 	

\section{Discussion}
\subsection{Origin of the planar defects}
\label{sec:discussion}

It is tempting to discuss the possible origin of the planar defects. 
	The physical approaches applied above (TEM, and DFT calculations) provide only the characteristics of the existing planar dfefects - their lattice structure, elemental composition,  formula, band structure and magnetization. However,  they don't answer the question on the mechanisms responsible for the planar defects emergence.

	Although we did not study this issue in this work, we would like to comment on it using a physicochemical approach:\\
	(i) Tin is known to manifest various valence states, from  (+2) to (+4), with comparable probabilities. \\
	(ii) The formation enthalpy  of the two competing phases (``122'' and ``112'')  are rather close (see Fig.~\ref{fig:TEM}h, Fig.~8 of  SM \cite{SM}, and  also Refs.~\cite{enthalpy,materials database}), providing a comparable probability of the emergence of the two phases during the high temperature  synthesis 	at about 850\,$^\circ$C.  
	Indeed, the	difference of enthalpy of the ``112''-phase and of the equilibrium mixture of the components and precursor:
	$$
	\Delta H = \Delta H_{112}-\Delta H_{\rm mixture} = 0.319{\rm eV/at.}
	$$
{\cb
Therefore,  the equilibrium defect concentration $C_{\rm def}$, estimated as probability of the ``112'' phase formation at $T=1123$K
	is   $C_{\rm def} = \exp(-\Delta H/kT)  =3.7\%$.
	(For more detail, see Section VI of SM \cite{SM}).
}
\subsection{Other type of defects and alternative origin of the ferromagnetism}

	Our data strongly suggest that planar defects are the source of the observed {\cb ferrimagnetic} properties of the AFM crystal. Alternatively, it is  very unlikely the  FM-type behavior  in  ESA	to appear from magnetic  impurities (as is often presumed in the literature), since the compounds are commonly synthesized from high purity raw materials. Rather, the ferromagnetic-type signal in pure compounds most likely originates from structural defects. In the studied 
	ESA 	compound the major defects are planar, whereas in other materials the structural defects may be different.
	Recently, in MnBi$_2$Se$_4$,  magnetic moments have been detected which are  induced by misplaced Mn atoms \cite{fukushima_PRM_2024}.  Furthermore, in Ref.~\cite{chen_PRB_2024} it was found that 5\% of Eu vacancies in EuZn$_2$P$_2$ causes ferromagnetic ordering.
	In contrast, in EuSn$_2$As$_2$, our EDX and  XRD data (see sections II and IV of SM \cite{SM}) doesn't reveal  Eu deficiency within $0.7\pm 0.3\%$, thus leaving the planar defects the most likely source of the {\cb ferromagnetic-type} features.

\subsection{ DC magnetization features and planar defects}
As we  mentioned above, in several publications, one can notice in low-fields a weak nonilearity of the  $M(H)$ dependences when magnetic field lies in the $ab$ plane. {\cb For the reader's convenience, we presented an exaggerated example 
	in Figs.~\ref{fig:M(H)}. Although more typical $M(H)$ curves (such as those shown in Fig.~9 of SM) appear virtually linear at first glance, zooming in on the low-field region reveals their non-linear nature in the range $\pm 0.1$T for $H \| (ab)$.}
It would be tempting to link the hysteresis to the same weak {\cb ferromagnetic-type} magnetization of the defects. Indeed, 
the fact that the non-linearity and hysteresis are observed only when $\mathbf{H}$ lies in the 
$(ab)$- plane is  natural for the easy-plane AFM ordering.

\subsection{On the potential coupling between  planar defects} 
Planar defects, considered as {\cb ``thin'' ferromagnetic (or ``thick" ferrimagnetic)} planes, are located at a significant distance 
from each other,  $d \sim 10-100$nm, and therefore practically do not interact with each other.
This fact follows both from simple theoretical estimates and from our measurements. 

On the theory side, using the text book one can  estimate the  magnetic field energy, and   exchange coupling $J_{\rm FM}$ for various 
mechanisms of interaction between the planar defects 
and make sure that coupling between remote planar defects is negligibly weak as compared with characteristic 
energy $k_B \Theta_{\rm FM}\approx k_B T_N$.

On the experimental side, this conclusion stems, e.g., from the  measured saturation magnetization. 
Firstly,  on the $M(H)$  curve (Fig.~\ref{fig:M(H)}b) the  FM-hysteresis saturation 
magnetization $M_{\rm s}^{\rm FM}$ is by a  factor of 37 less than the saturation magnetization $M_s$ of the regular AFM crystal 
(c.f. Figs.~\ref{fig:M(H)}a and \ref{fig:M(H)}b). Given the defects on average occupy about 3\% of the sample volume, we conclude that the measured saturation FM magnetization $M^{\rm FM}_s$ does not exceed the anticipated magnetization of the sum of individual defect layers, $3\%M_s$; this means that the defects respond individually, rather than collectively as in a bulk ferromagnet.

{\cb
Secondly, the ferromagnetic contribution to DC magnetization (Fig.~\ref{fig:M(H)}b) saturates in field $H_s^{\rm FM} \sim 30$mT that is a factor of 120 lower than the AFM saturation field $H_s \approx 3.6$T (in plane), see Fig.~\ref{fig:M(H)}a. Since the exchange field  $H_e$ roughly equals  $H_s^{\rm FM}/2$, we conclude that the  defect-defect exchange energy $J \sim  H_e M$ is by a factor of $\sim 120\times 37=4\times 10^3$  lower than the usual AFM interplane exchange between the neighboring Eu layers 
$J_{\rm AFM}=+0.19$meV  \cite{sun_SciChina_2021, grigorev-CommunMater_2025}.
Similarly, the magnetic dipole-dipole interaction energy between defects located at an average  distance of $r\sim 24$\,nm   decays with distance $r$ as $(1/r)^3$ and is negligibly small. 
The stray field, therefore, is negligibly small and the defects do not ``feel'' each other.
Thus,  the above estimates and DC magnetization measurements show that the FM-moments of planar defects respond individually  (i.e. as a set of independent ferromagnetic pieces), rather than collectively as a single bulk ferromagnet. 
}

	 \subsection{On the spatial orientation of the 	{\cb magnetic} vector of the defect.} 
\begin{figure}[ht]
\includegraphics[width=120pt]{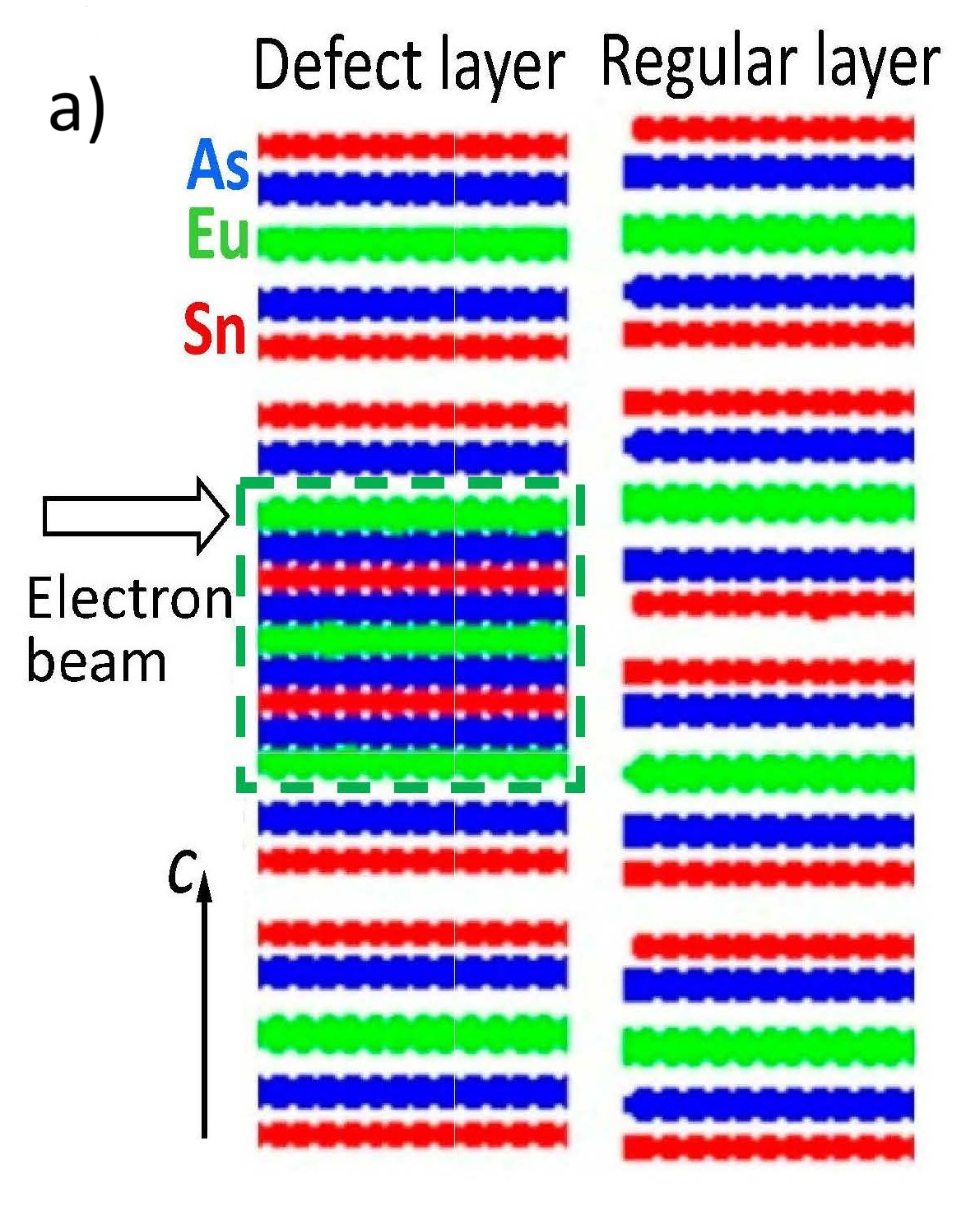}
\includegraphics[width=100pt]{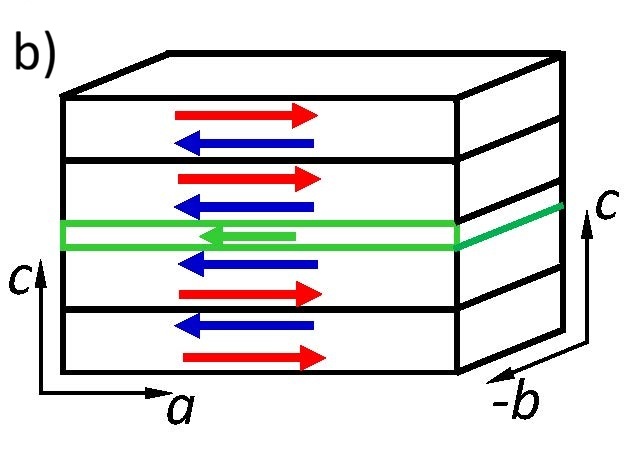}
\caption{{\cb (a) Schematic model  of the  specimen cross section with   defect layer (enclosed in a green frame)  and regular  layers. (b) Schematic model of the  arrangements of Eu moments in the defect layer and the bulk crystal. 	 Blue and red arrows depict magnetization vectors of the Eu-sublattices in the regular layers, and  green arrow - in the defect layer. NB: the actual angle $\angle ab =120^{\circ}$.  For more detail, see Fig.~9  of SM.
	}
} 
\label{fig:spatial arrangement}
\end{figure}

		An interesting question concerns the orientation of the magnetization of defects. No doubt, the defect magnetization vectors  $\mathbf{M}^{\rm FM}$ at zero field 		lie in the $(ab)$ plane. Regarding orientation of the $\mathbf{M}^{\rm FM}$-vectors relative to each other, 		we noted above that the interaction  energy between the defects is negligibly low and hence, the orientation 
		of the $\mathbf{M}^{\rm FM}$ 	is determined primarily by the interaction of the defective layer with adjacent lattice layers.
	{\cb Our DFT calculations explicitly indicate their parallel (i.e., ferromagnetic-type) mutual orientation.
	Having no direct experimental data 		(such as scanning magnetometry with atomic resolution) to test this prediction,
		we depicted in Fig.~\ref{fig:spatial arrangement}b the most likely possible configuration, and present below and in section VIII of SM some arguments in its support.		
		If the defect extended across the entire width (length) of the bulk specimen, from its edge to edge in $(ab)$ plane, than 
	the $\mathbf{M}^{\rm FM}$ orientation would be governed by interlayer exchange interaction between the defect layer and the  neighbor atomic layers of the supercell; the latter is of ferromagnetic sign as follows from DFT calculations (see section \ref{sec:DFT}). 
	This case is illustrated  in Fig.~\ref{fig:spatial arrangement}b.

		However, even on the limited area of lamella, we observed  points of defect termination at a dislocation (see  Fig.~7 of SM \cite{SM}). In such cases, we can't exclude that the 	preferred  orientation of magnetic moments (Fig. ~9b) will be violated.
The general situation then becomes somewhat similar to that known in ferrimagnets 	\cite{chikazumi} where upon lowering temperature, two FM-ordering transitions occur. Applying this line of reasoning to our samples, we propose that the higher-temperature ferromagnetic transition at 24\,K,  and yet another one in the interval    16-19K, may correspond to various magnetization arrangements.
}

\subsection{On the origin of FM-hystereis}		
 The hysteresis of the $M(H)$ curves in Fig.~\ref{fig:M(H)}b may be caused by a weak anisotropy, i.e. a nonzero constant $K_u^{(ab)}$, as mentioned above.
In zero external field, the FM-moments $\bf{M}^{\rm FM}$ should be directed equally likely in  possible directions $a$ and $b$ in various defect layers depending on the magnetization orientation in the neighboring regular Eu-layers. 
In addition, due to the finite length of defects (see Section V of SM \cite{SM}), FM-domains and domain walls must necessarily exist.
The hysteresis width, apparently, is a consequence of the finite energy required to reorient part of the FM-moments along the field  direction. 
Recently, a weak anisotropy in the $ab$ plane of EuSn$_2$As$_2$ was observed below $T_N$  from measurements of the second harmonic generation \cite{saatjian-2407.03459}, consistent with the above  discussion.

The low-field ferromagnetic type $M(H)$ hysteresis and nonlinearity (Figs.~\ref{fig:M(H)}a,b) disappear at a temperature of $\approx 24$\,K (Fig.~\ref{fig:M(H)}c), that  almost coincides with the Neel temperature $T_N=24$K of the AFM ordering in the bulk, determined from disappearance of nonlinearity in the antiferromagnetic $M(H)$ characteristics in higher fields $\sim 4$T 
\cite{golov_JMMM_2022}. 	The latter coincidence points at a link of the hysteresis with the AFM ordering in the bulk on average, in addition to the  link with local FM-ordering in the defect areas (which seems to   set at $T_c=18-19$\,K, as follows from the AC-susceptibility (Fig.~\ref{fig:chi(T)}). As noted above, the magnetic ordering process in the temperature interval of 18–24\,K resembles the process characteristic of ferrimagnets.

\subsection{On the AC magnetic susceptibility}
We mentioned in section \ref{introduction} and showed in Fig.~\ref{fig:chi(T)}, that the low-frequency AC susceptibility $\chi(T)$  measured for EuSn$_2$As$_2$ crystals in low fields often exhibits an upturn, below $T_N$.  
This upturn in Fig.~\ref{fig:chi(T)} starts somewhat lower than the Neel temperature $T_N\approx 24$\,K for the bulk crystal. In some papers, this divergence is not seen simply because it is washed out by applied field of $H>0.05$\,T which polarizes the  FM-moments.   

The $\chi(T)$ divergence points at 	a potential ferromagnetic transition developing in minority fractions of the bulk crystal  that already has antiferromagnetic ordering. We consider this divergence as a {\em direct}  manifestation of ferromagnetic ordering in the minority fraction of the crystal (the planar defects)  at $\Theta_{\rm FM}= 18-19$\,K. 
 Within our approach, the observation of the weak FM-type hysteresis in low fields $<0.01$\,T  {\cb is consistent} with
 the presence of the FM-defects and explains both, the $M(H)$ nonilearity and the $\chi(T)$ divergence.

	\subsection{Manifestation of magnetic defects in other physical properties of the vdW crystals}
		The FM-defects are expected to pin the magnons. However, 
		experimental verification of the presence of magnon-gap  is challenging: testing the potential magnon gap and the hysteresis spin-canting 
	by electron  spin resonance (ESR) measurements \cite{golov_JMMM_2022}  
	in such low fields ($H< 0.03$T) would  require lowering  microwave frequency by  a factor of 10,
	that would drastically decrease the Zeeman energy  and impede ESR observation.
		Nevertheless, we believe,	the ferromagnetic planar defects are  rather likely to be 
	responsible for the splitting of the ESR line observed in Ref.~\cite{golov_JMMM_2022}.

\subsection{On the possibility of detecting  the defects and their magnetic structure by other techniques}
The best technique to visualize the ferromagnetic planar defects is the scanning SQUID or NV-centers scanning microscope; it would enable to detect local ferromagnetic moments of the ``odd'' Eu atoms and to visulalize the spatial magnetic structure.
	In view of the small concentration of planar defects in the sample, the presence of additional bands and the energy shift of the density of states for some  bands 	is unlikely to be visible in XPS and ARPES on top of 	the background of the intensity from other bands and the density of states of atoms from the regular ordered structure. 
		Regarding 	magnetotransport measurements,  the in-plane charge transport is insensitive to the presence of low density planar defects whose thickness is less than the mean free path; 
	in experiments  \cite{pervakov_PRB_2026}, the magnetoresistance  was found to be  isotropic in the $ab$-plane.
	
	\subsection{ Potential applications of the  metamaterial}	
The EuSn$_2$As$_2$ crystal with imbedded planar defects may be considered as a metamaterial. By applying rather small magnetic field ($\sim 100$\,Gs), the subsystem of defects can be switched from the AFM to FM state, without affecting the large AFM magnetization of the host lattice. Due to this property, 	EuSn$_2$As$_2$ and similar vdW layered AFM semimetals might find an application in spintronics. 

\section{Conclusion}
\label{sec:conclusions}

To conclude, in this work we investigated high purity EuSn$_2$As$_2$ single crystals which experience antiferromagnetic ordering  at $T<T_N$. This compound exhibits several unusual magnetic properties which remained puzzling by now. These pecularities include magnetic susceptinbility divergence at $T\rightarrow 0$, in-plane magnetization nonlinearity in weak fields, and splitting the ESR line \cite{golov_JMMM_2022}. 
 In search of the origin of these features we performed precise magnetization measurements and found a weak ferromagnetic type hysteresis in low fields. The hysteresis is evidently the cause of the magnetization nonlinearity. This indicates the presence of intrinsic ferromagnetic inclusions in the high-purity stoichiometric compound, yet it does not explain their origin. 
To determine the source of ferromagnetism we applied a combination of EDX, HRTEM, and EELS techniques, which revealed the existence of nm-thick (along the $c$-axis)  and several hundred nm long (in a perpendicular direction) planar defects; the defects  aligned  perpendicular to the $c$ axis are randomly distributed in the bulk. 

Using the  TEM investigateions, structural analysis,  and DFT calculations  we identified  the chemical formula of the local phase in the planar defect area to be EuSnAs$_{2}$.
Incorporating the cubic defect lattice structure into the rhombohedral defect-free bulk lattice 
requires the defect unit cell to comprise 7 Eu atoms with resulting composition of the  transition defect supercell  Eu$_7$Sn$_{12}$As$_{14}$. Due to the odd number  of Eu atoms, the defect 
posesses the intrinsic non-zero magnetic moment,  
that is also promoted by the locally enhanced exchange interaction between Eu atoms. 

From general arguments, the moments are expected to lie in the easy magnetization $ab$-plane of the host crystal and are oriented along the AFM vector $\mathbf{L}$. 
Upon cooling below $\approx 24-18$\,K,  the magnetic moments of the defect layer experience ferromagnetic-type 
ordering  which may  explain the observed  {\cb ferrimagnetic-type} magnetization of the AFM crystal. 
In order to complement and to substantiate the structural finding
we performed extensive DFT calculations of the defect band structure, energy levels, density of states and magnetization. Their results, in particular,  confirm the weak in-plane ferromagnetic moment of the defect, $\approx 0.2\mu_B/$f.u. 

This theoretical conclusion was supported by our high-precision magnetization measurements in weak magnetic fields, which revealed a ferromagnetic-type hysteresis. We suggested,  the hysteresis might originate from  an in-plane magnetic anisotropy caused by extended defects, i.e. from  in-plane pinning of the magnetization by the anisotropic magnetic defects.
 We believe, the extended defects spontaneously formed in the bulk antiferromagnetic crystal are responsible
for the unusual magnetic and  microwave resonance properties of the EuSn$_2$As$_2$ single crystals.  

{\cb
Our results  
demonstrate the existence of a natural metamaterial composed of a bulk antiferromagnetically ordered EuSn$_2$As$_2$ matrix and ferromagnetic EuSnAs$_2$ monolayer inclusions. 
This finding sheds new light on the mechanism  of ferromagnetism in quasi-two-dimensional antiferromagnetic materials 
and may find practical application in the creation or design of natural interfaces between antiferromagnetic and ferromagnetic phases.
}

\section{Contributions}
A.Yu.L., A.L.V., A.V.O., and V.I.B. performed the STEM,  EDXS, and EELS measurements, and processed their results. 
N.S.P. performed the DFT calculations.  
K.S.P. and V.A.V. grew and characterized the crystals. A.Yu.L., A.L.V., V.I.B., and K.S.P. analyzed the defect lattice structure.
A.V.S. performed magnetic measurements, 
V.M.P., A.Yu.L. and A.V.L. analyzed the data and wrote the paper. V.M.P. conceived this project. All the authors discussed the results and offered useful inputs.

\section{Declaration of Competing Interest} 
The authors declare no conflict of interest.

\section{Data availability}
All the data 
are available  from the corresponding author upon a reasonable request.

\section*{Acknowledgments}
AYuL, AVS, KSP, VAV,  and VMP acknowledge the support of the State assignment of the Ministry of Science and Higher Education of the Russian Federation (Project No. 0023-2019-0005). 
VMP is greatful to E. Deiatov and N. Orlova for useful discussion.
TEM measurements were performed using the equipment of the resource center of the NRC ``Kurchatov Institute''.
Synthesis, magnetic and transport measurements were performed using research equipment of the LPI shared facility center.

\newpage

\section{Supplemental Material\\ to the paper
	``Emergence of Ferromagnetism from Planar Defects in EuSn$_2$As$_2$ Antiferromagnet''}

	\author{A. Yu. Levakhova}
	{\email{levakhovaayu@lebedev.ru}}
	\author{A. L. Vasiliev}
	\author{N. S. Pavlov}
	\author{A. V. Ovcharov}
	\author{V. I. Bondarenko}
	\author{A. V. Sadakov}
	\author{K. S. Pervakov}
	\author{V. A. Vlasenko}
	\author{V. M. Pudalov}
	\date{\today{}}
	\maketitle
	
	\section{Samples for TEM studies}
	\label{SM:samples}
	
\begin{figure}[h]
		\includegraphics[width=0.37\textwidth]{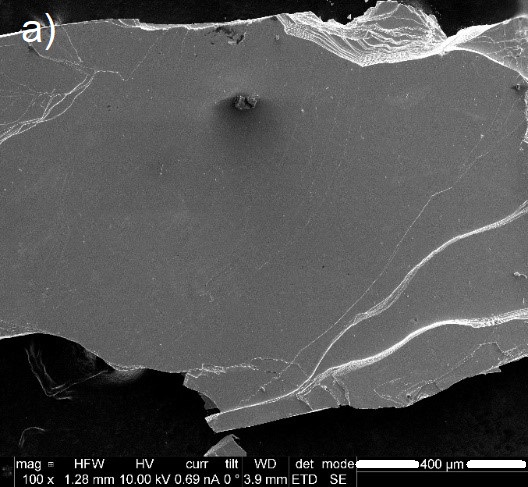}
		\includegraphics[width=0.37\textwidth]{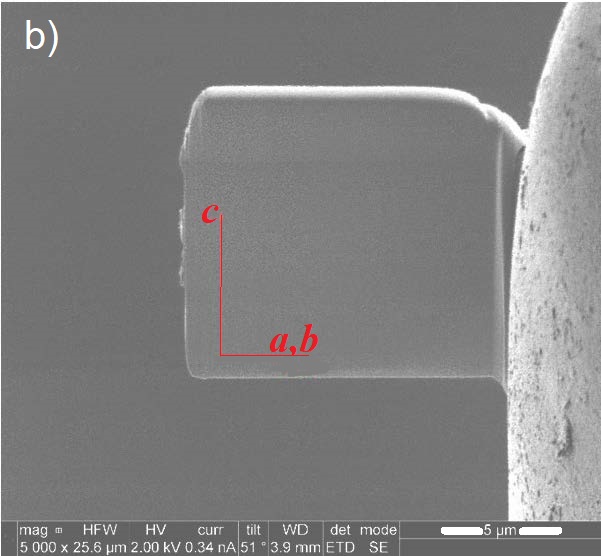}
		\caption{(a)Image of a crystal with a flat face ($ab$-plane). White bar shows the  400$\mu m$ length scale; (b) SEM image of the 15\,nm thick  lamella cut from the crystal. White bar shows the $5\mu m$ length scale.	} 
		\label{fig:SM1}
	\end{figure}

	Figures \ref{fig:SM1}a,b show  a scanning electron microscope  image  of the single crystal  selected for slicing, and of the lamella, $15\times 15\mu$m  cut from the bulk sample along the  $ac$ crystal plane.
	
	\newpage
	\section{Elemental composition}
\begin{figure}[h]
		\includegraphics[width=250pt]{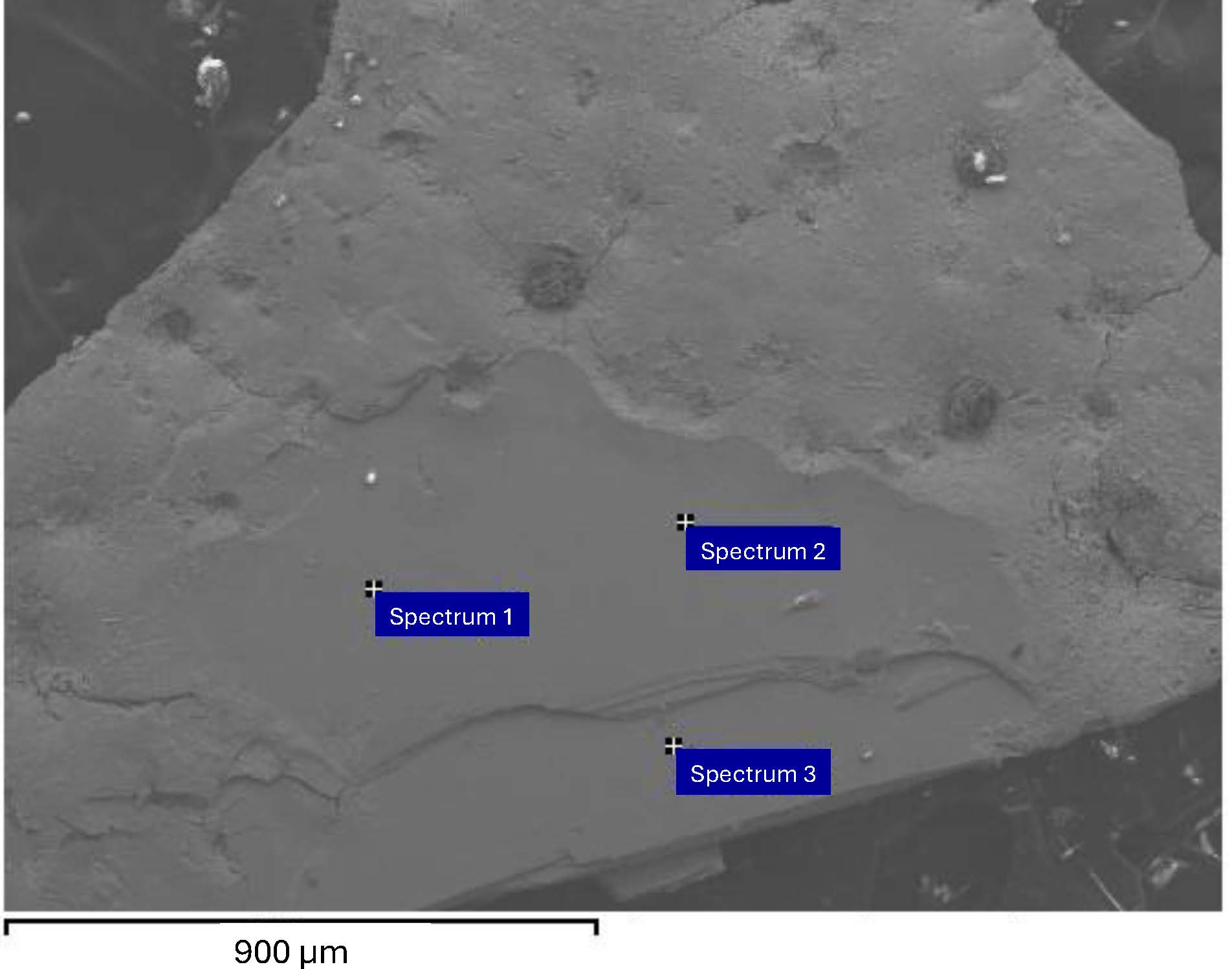}
		\caption{ The crosses mark three spots at the sample surface, where elemental composition was measured by EDX technique.}
		\label{fig:SM-EDX}
	\end{figure}

	EDX analysis was done using JEOL JSM 7001F scanning electron microscope equipped with INCA X-act analyzer.
	The obtained data were processed using the INCA software package.
	Figure ~\ref{fig:SM-EDX} shows SEM image  of a typical bulk sample from the same batch. The crosses mark  three positions  at the surface of the sample where the analysis was taken. The results  are summarized in Table \ref{table:0}.

	\begin{table}[h]
		\caption{Results of the EDX  analysis (in at. \%) }
		\begin{tabular}{|c|c|c|c|c|}
			\hline
			Elements	&	As &	Sn &	Eu &	Total\\
			\hline
			Spot  (1) &	41.90 & 38.29 & 19.81 & 100.00\\
			\hline
			Spot (2)  &	41.49 & 38.63 & 19.88 & 100.00\\
			\hline
			Spot (3) &	41.78 & 38.33 & 19.89 & 100.00\\
			\hline
			Average	& 41.72	& 38.42	& 19.86	& 100.00\\
			\hline
			$\sigma$  &	0.18	&0.21 &	0.05	&   \\
			\hline
			Formula coeff.	& 2.086(4)	& 1.921(5)	&0.993(3) &	\\
			\hline
		\end{tabular}
		\label{table:0}
	\end{table}
	
	As can be seen from the table, within the measurement error, there is virtually no deficit of Eu atoms; more precisely, the Eu deficit is $0.7\pm 0.3\%$.
	The overall deficit of Sn-atoms is $3.95\pm 0.25\%$.
	
	\section{X-ray crystal diffraction I}
	\label{SM:XRD}
\begin{figure}[h]
		\includegraphics[width=0.5\textwidth]{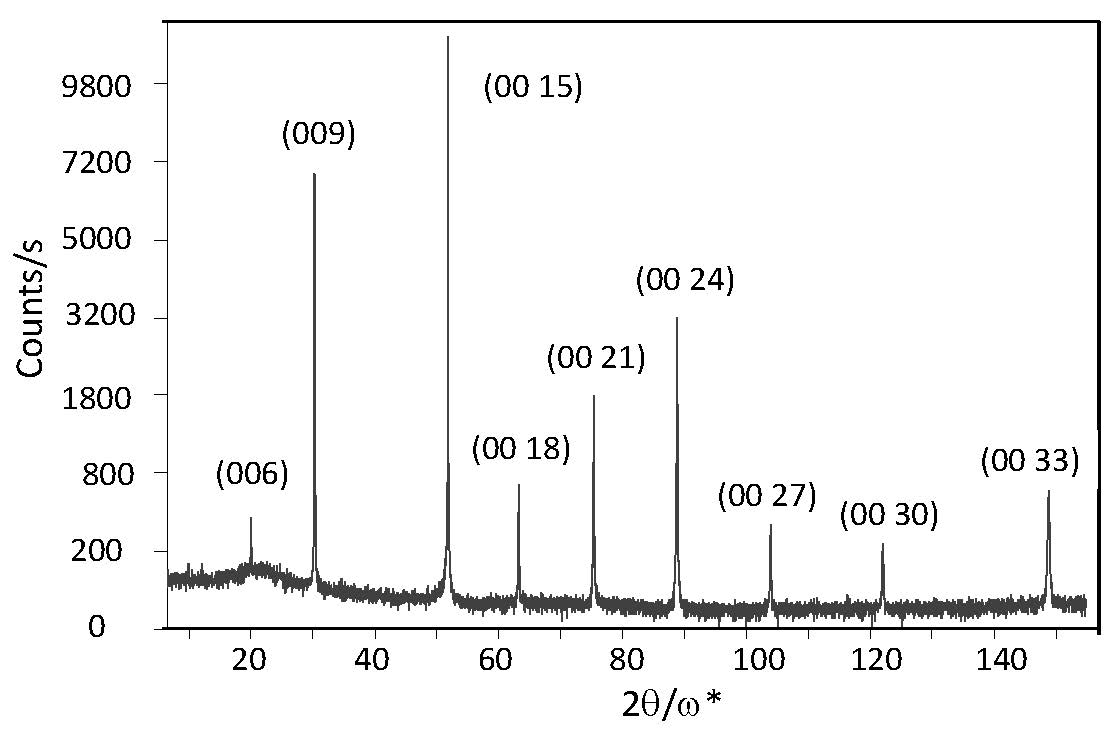}
		\caption{X-ray diffraction pattern on a square scale for the studied EuSn$_2$As$_2$ bulk crystal 
			from the side of the shiny surface. 
			Correction for hybrid monochromator $\Delta(2\theta)=0.0135^\circ$.} 
		\label{fig:SM2}
	\end{figure}

	In order to test crystal structure,  measure lattice parameters and evaluate  lattice	defects we performed precise XRD examination of the bulk sample -
	with PanAlytical X’pert Pro X-ray diffractometer equipped with multiple crystal analyzers.
	Figure \ref{fig:SM2} shows X-ray diffraction pattern on a square scale for the  EuSn$_2$As$_2$ bulk crystal.
	One can see the presence of all possible peaks of the $(00 l)$ series for a 
	rhombohedral EuSn$_2$As$_2$ lattice ($l=3n$, where $n$ is an integer),
	as well as the absence of a noticeable broadening of the farthest reflections; these results are the evidence of a macroscopic structural perfection of the sample in the direction perpendicular to the basal plane. 		
	
	\begin{figure*}[h]
		\includegraphics[width=200pt]{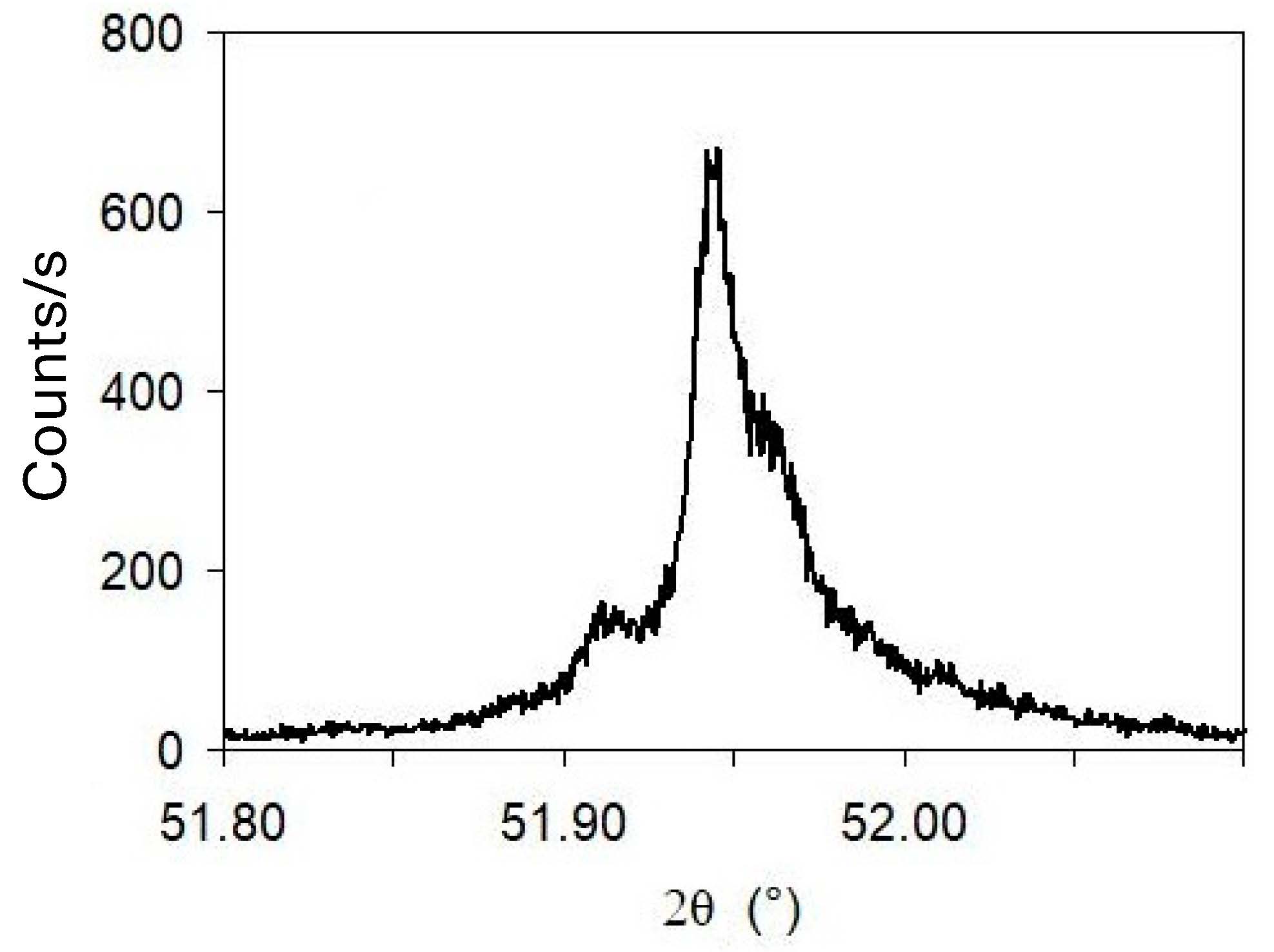}
		\includegraphics[width=200pt]{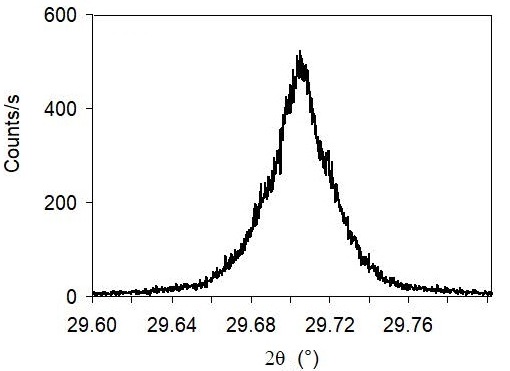}
		\caption{
			(a) $2\theta -\omega$ scanning curve measured with the third analyzer crystal. 
			(b)	$ 2\theta - \omega$ scanning curve with the third crystal-analyzer on the asymmetric reflection (10\,5) in azimuthal position $\varphi=12^\circ$. 
		}
		\label{fig:XRD2}
	\end{figure*}
	
	Figure \ref{fig:XRD2}a
	measured with the 3rd analyzer crystal  reveals a small spread in the lattice parameter for individual layers along the $c$ axis. For the main maximum we find $2\theta = 51.9436^\circ$, and taking into account the correction for the hybrid monochromator, we eventually, obtain $c =26.3780\AA$.
	Figure  \ref{fig:XRD2}b	shows $(2\theta - \omega)$-scanning curve with the third crystal-analyzer on the asymmetric reflection (10\,5) in azimuthal position $\varphi=12^\circ$.
	Note that all the curves without exception were measured under full illumination of the sample with X-ray beam.  Based on the angular position of $2\theta_{(10 5)}=29.7054^\circ$ in Fig.~\ref{fig:XRD2}b,  taking into account the correction for the hybrid monochromator, and the known value $c=26.3780$\AA, we find $a_{(10\,5)}=4.2190$\AA. 
	
	Comparing with the literature data ($c=26.463 \AA$, $a=4.2071\AA$), we conclude that in the studied crystal, lattice parameter along the $c$ axis is lowered by 0.3\%, and along the $a$ axis  is increased. 
	The  smaller $c$-axis spacing might be related with the presence of planar defects in which the  Eu-Eu spacing is smaller.
	However, the 0.3\% difference  is by a factor of 3 .7   smaller than the anticipated 1.1\% compression due to the presence in the bulk of 3\%   defect layers (each thinner by 0.5nm).  This inconsistency may indicate that the defects extend not through the entire bulk sample along the $(ab)$-plane.
	
	\section{X-ray crystal diffraction II}
	In order to doublecheck the results, and to measure lattice site occupancies 
	we performed an independent XRD examination using Tongda TD-5000  single crystal diffractometer, equipped with CCD detector 
	and Mo - K$\alpha$ microfocus X-ray  tube. For absorption correction we used SCALE3 ABSPACK [ABSPACK, S. (2012). Empirical absorption correction. CrysalisPro–Software Package Agilent Technologies].
	
	\begin{figure*}[ht]
		\includegraphics[width=150pt]{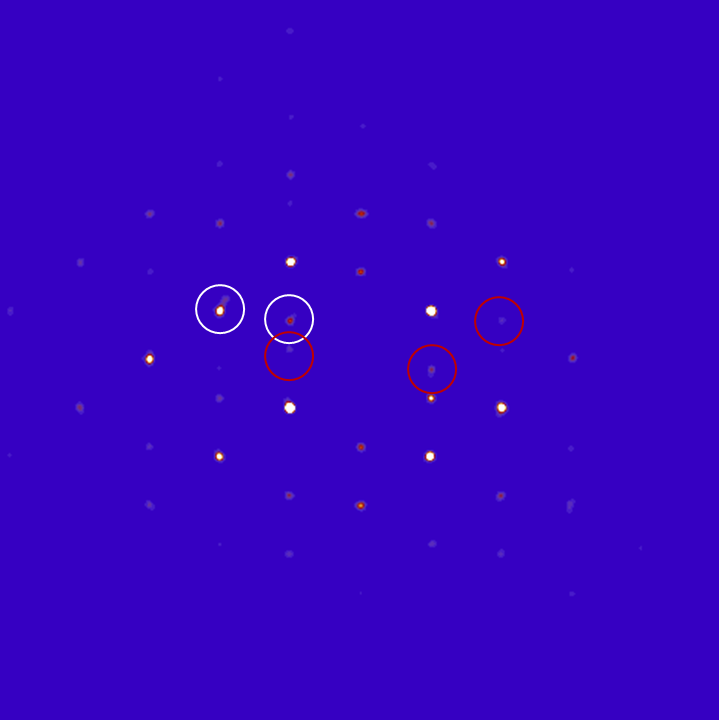}
		\includegraphics[width=150pt]{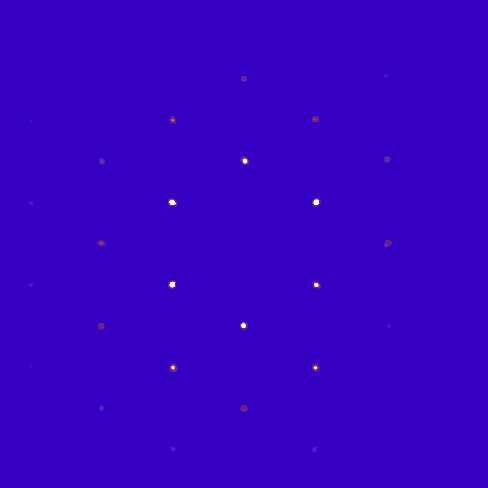}
		\includegraphics[width=150pt]{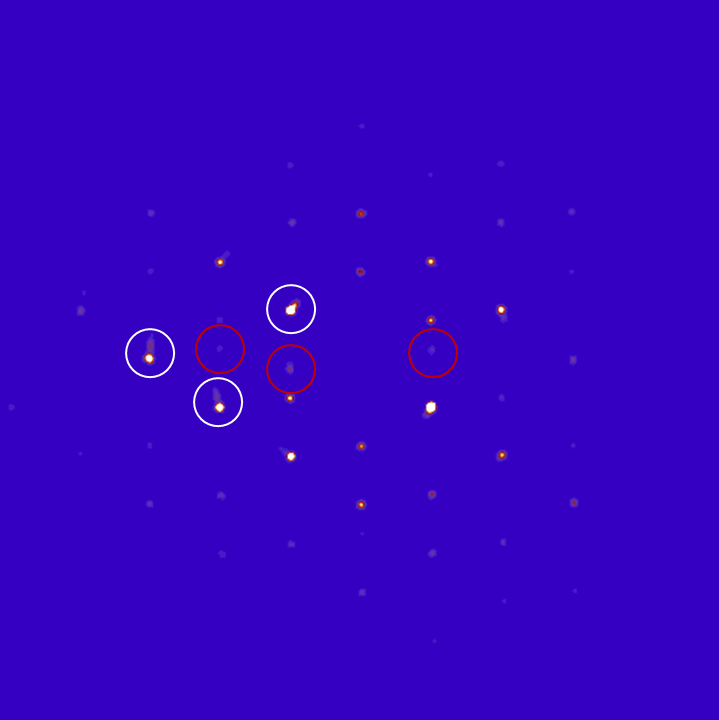}
		\caption{Diffraction patterns (a) in [h00]-, (b) in [00l]- , and (c) in [0k0]-  directions. In panels (a) and (c), reflections corresponding to layer twisting are marked with white circles, and reflections corresponding to tin atom displacement are marked with red circles.}
		\label{fig:Tongda}
	\end{figure*}
	
	Lattice  structure was solved using SHELXT-2018 [Sheldrick, G. M. (2015). 
	Foundations of Crystallography, 71(1), 3-8.] program and refined with SHELXL-2018 [Sheldrick, G. M. (2015). 
	Crystal  Structure Commun., 71(1), 3-8.] program included in OLEX2 software [Dolomanov, et al.
	Appl. Crystallography,  {\bf 42}(2), 339-341 (2009)]. 
	
	Diffraction pattern in   [00l]  direction (Fig. \ref{fig:Tongda}) shows clear diffraction spots corresponding to high crystalline quality of the sample in $ab$ plane. However, the diffraction patterns in the [0k0] and [h00] directions show evidence of slight twisting of the layers along the $c$ axis, resulting in blurred diffraction spots around the central point.
	
\begin{figure*}[ht]
		\includegraphics[width=150pt]{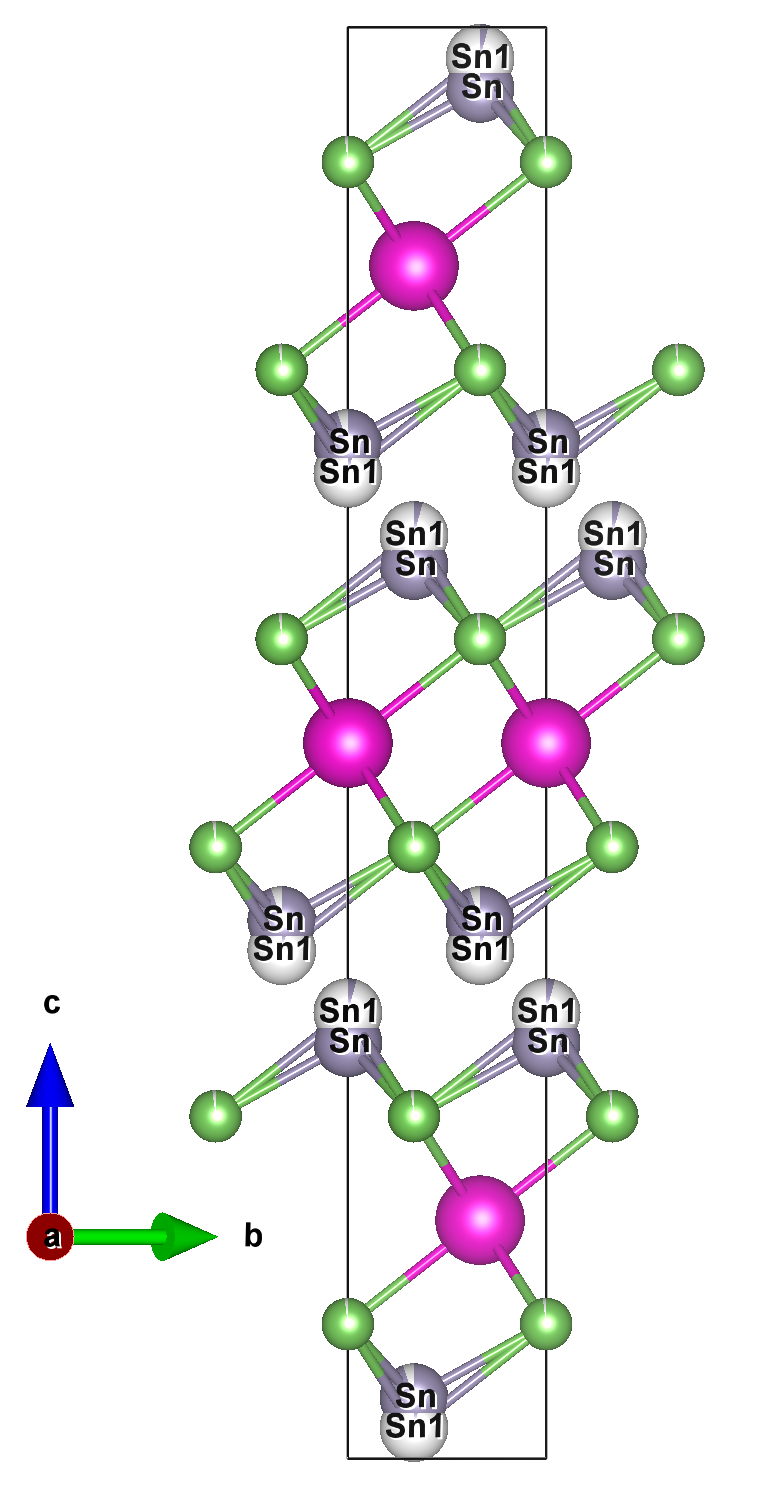}
		\caption{Two Sn-positions in the elementary cell, as follows from the  single crystal X-ray diffraction experiment.}
		\label{fig:two Sn-positions}
	\end{figure*}

	Crystal data,  details of experiments, and crystal structure refinement are given in Table \ref{table:1}. Refined atom positions, occupancies and displacement are listed in Table \ref{table:2}. Planar defect inclusions of 112-type in crystals could be a reason for splitting of Sn position into two separate ones along the $c$-axis, as schematically illustrated in Fig.~\ref{fig:two Sn-positions}.  Conclusion on the splitting of atom positions follows from the observed  high electron density peak near the Sn-atom position, which has been confirmed with many crystals. Distance along the $c$-axis between the two Sn-positions is 0.55(11)\AA  (i.e., the differences in W-Z cell positions multiplied by the cell parameter $c=26,33\AA$ in Table \ref{table:2}). Accounting for this effect also leads to significant increase of the structure refinement quality. 
	
	According to our XRD data, the obtained occupancy in the Sn1 position corresponds to the tin in defect layers and is in a good agreement with the TEM data in part of defects volume fraction.
	Noteworthy, the shift of the Sn1 position off the surface along the $c$-axis is also consistent with 
	TEM spectra of the defect layer in Fig. 5 of the main text.
	
	\begin{table}
		\caption{Crystal data collection and refinement parameters for EuSn$_2$As$_2$}
		\begin{tabular}{|c|c|}
			\hline
			Refined chemical formula & EuSn$_{1.97}$As$_{1.96}$\\
			\hline
			M$_r$ & 531.98\\
			\hline
			Crystal system, space group &  Trigonal, $R-3m$ \\ 
			\hline
			Temperature (K)	&  293(2)\\
			\hline
			$a, c$ (\AA)	& 4.2113(9), 26.329(5)\\
			\hline
			$V$ (\AA$^3$)	& 404.38(19)\\
			\hline
			$Z$  &	3\\
			\hline
			Radiation type &	Mo K$_\alpha$\\
			\hline
			$\mu$ (mm$^{?1}$)	& 32.272\\
			\hline
			Crystal size (mm)	& $0.08\times 0.06\times 0.03$\\
			\hline
			Diffractometer	& Tongda TD-5000\\
			\hline
			Absorption correction &	ABSPACK\\
			\hline
			$R_{int}$  &	0.1361\\
			\hline
			$R[F^2 > 2\sigma(F^2)]$, $wR(F^2)$, $S$ &	0.0416, \.0.0984, \.1.143\\
			\hline
			No. of parameters	& 13\\
			\hline
			$\Delta\rho_{max}, \Delta\rho_{min}$ ($e\AA^{-3}$)&	3.23, -1.18\\
			\hline
		\end{tabular}
		\label{table:1}
	\end{table}

	We note that the lattice parameters determined in two independent  XRD experiments with bulk crystals from the same batch  (section  II, and here,  Table II) are slightly different. We believe this is due to  manual determinations of the peak positions in the first case  that  didn't account for   non-Gaussian shape of the peak. 
	
	More detailed information about filling the crystal lattice sites is presented in Table \ref{table:2}.
	\begin{table}
		\caption{Refined atomic coordinates in the Wigner-Seitz-cell, and position occupancies }
		\begin{tabular}{|c|c|c|c|c|}
			\hline
			Atom	 &x         &	y	   & z	                   & Occupancy\\
			\hline
			Eu	        & 1	       & 0        &0.5                  & 1\\		
			\hline
			Sn (2+)	 &0.333	&0.667 &	0.3760(2)  &	0.945(17)\\
			\hline
			Sn1 (4+)&0.333	&0.667 &0.355(4)	   & 0.038(15)\\
			\hline
			As	        &0.667	 &0.333	&0.42738(9)	 &0.979(12)\\
			\hline
		\end{tabular}
		\label{table:2}
	\end{table}
	As follows from the Table, 
	the total Sn  deficit  is $1.7\pm 3.2\%$ and doesn't contradict the results of EDX measurements. 
	In other words,  the   total  amount of Sn atoms consists of the two components, $94.5\pm 1.7 \%$ of  Sn(2+) and   $3.8 \pm 1.5\%$ of Sn(4+).
	The   amount of Sn1(4+) is consistent with the $\sim 3\%$ volume fraction of the planar defects (see below).

	\newpage
	\section{Transmission electron microscopy studies}
	\begin{figure}[h]
		\includegraphics[width=0.48\textwidth]{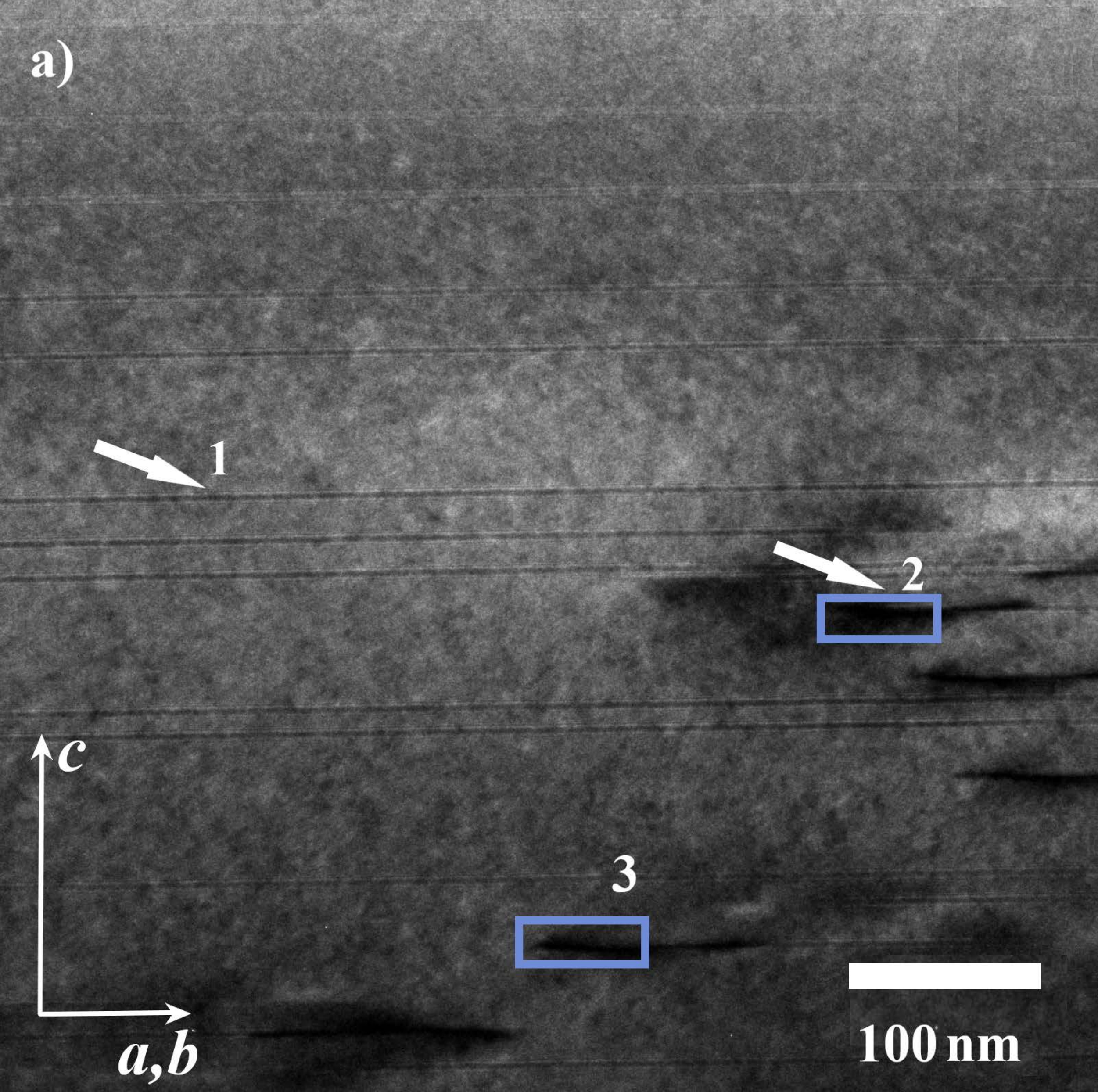}
		\includegraphics[width=0.48\textwidth]{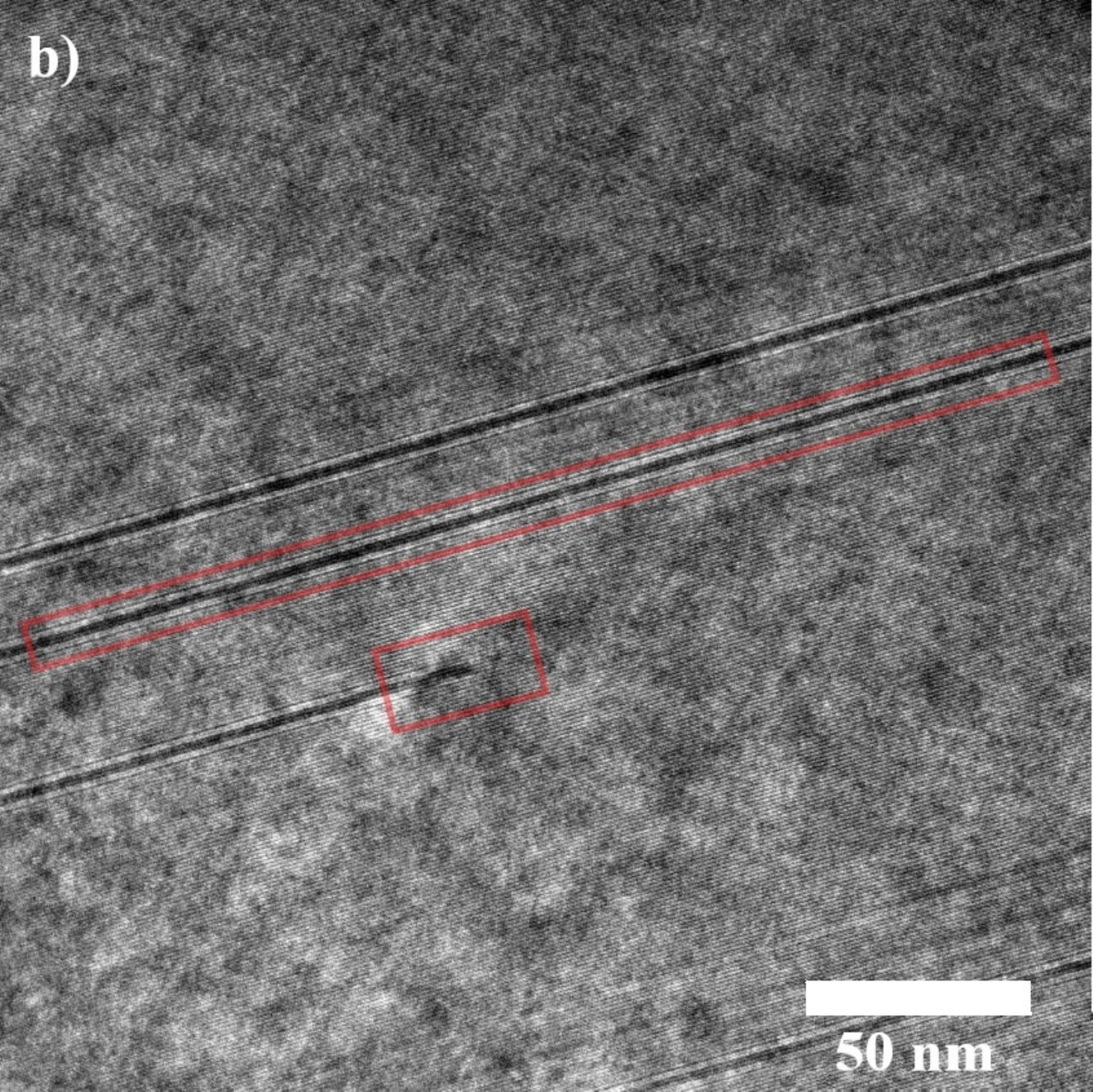}
		\caption{BF TEM images of  EuSn$_2$As$_2$ single crystal with planar defects;   
			white arrow \#1 on the panel (a) points at one of the planar defects. Blue rectangles frame the dislocations.
			(b) The upper red frame highlights one of the planar defects, the lower red frame encloses the end of the planar defect. White bars depict the length scale, 100\,nm  on panels $a$ and $b$.} 
		\label{TEM}
	\end{figure}
	
	{\cb 
		The proportion of volume occupied by defects was estimated by mapping the TEM image in the $ac$-plane.   
		Since the information on the defect size along the $b$ crystal direction (perpendicular to the lamella plane) was unavailable, 
		we presumed the defect width and length  (i.e. along $a$ and $b$ axes) are the same. Two typical high resolution TEM images of the $ac$-surface of the lamellae are shown in Fig.~\ref{TEM}. 
		By zooming in the image of the panel (a) one can notice 20 lines of planar defects, 12 of which extend over the whole field of view (580\,nm) and 8 defects end or begin within the
		inspection area (examples are enclosed in rectangles in Figs.~\ref{TEM}). Knowing the defect thickness, $0.8$\,nm,  we obtain the sought for a rough estimate  of 3.6\% volume occupied by defects. Similar mapping was done over several images, and resulted in the average  value $3 \pm 1\%$ which we take through the paper as an estimate of the  relative defect volume in the studied bulk crystals.
		
		An interesting question: is there any periodicity in the arrangement of planar defects along the $c$ axis?
		Looking at Fig. ~\ref{TEM}, one might get the impression that there is a certain local tendency for them to repeat 
		through a stack of $6n$ regular layers. 
		As an example,  within the field of view in Fig.~\ref{TEM}a the distances between 14 planar defects are as follows (in units of $c/2 =1.3$\,nm):
		5.9, 	 9.7, 	 11.5, 	 13.2, 	 17.2, 	 23.0, 	 30.9, 	 37.7, 	 38.0, 	 53.9, 	 55.7, 	 59.2, 	 72.5.    However, crystal XRD examination doesn't confirm the presence of such  superstructure on a macroscopic scale, and  we conclude that the defects are distributed rather randomly in the bulk. 
	}
	
	\newpage
	\section{Thermodynamics of the  phase formation in the  Eu-Sn-As system}
	
	\begin{figure}[h]
		\includegraphics[width=0.48\textwidth]{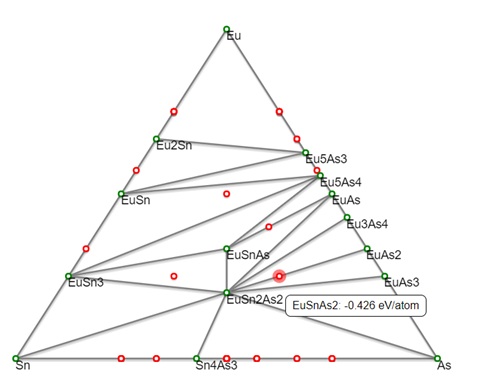}
		\caption{Convex hull showing formation enthalpy of various phases in the Eu-Sn-As system.}
		\label{figSM:convex hull}
	\end{figure}
	
	{\cb 
		Figure \ref{figSM:convex hull} shows ternary convex hull diagram of the Eu-Sn-As system. 
		[Wang, H.C., Botti, S. , Marques, M.A.L., Predicting stable crystalline compounds using chemical similarity. 
		npj Comput Mater {\bf  7}, 12 (2021). https://doi.org/10.1038/s41524-020-00481-6].
		
		For the two most closely located relevant ``122'' and ``112'' phases, the formation enthalpy  
		$\Delta H_{122}=-0.658$eV/atom,  and  $\Delta H_{112} =-0.426$eV/atom.
		In the synthesis reaction, the ``112'' phase is competing  with an equilibrium mixture of EuAs$_2$ and EuSn$_2$As$_2$, 
		for which $\Delta H_{\rm mix}=-0.745$eV/atom:
		\begin{center}EuAs$_2$ + EuSn$_2$As$_2$ $ \leftrightarrow$ 2 EuSnAs$_2$.\\
		\end{center}
	}
	For estimating the equilibrium content of the 112 phase in the mixture at  temperature $T=1123$\,K, one has to use the ``Hull distance'', 
	i.e. the the difference in enthalpies of the phase and the equilibrium composition:
	$$
	\Delta H= \Delta H_{112} - \Delta H_{\rm mix}=-0.426+ 0.745= 0.319 \textrm{ eV/atom}
	$$ 
	The equilibrium defect density taken as the probability of the (112)-phase formation 
	is therefore equals  $\exp(-\Delta H/k_BT) = 3.7\%$.
	
	This estimate is consistent with the concentration of defects in the volume, $3\pm 1$\%,  determined from TEM (Section V of SM).
	
	\newpage
	{\cb
		\section{High-field magnetization measurements}
		Because of the complex  structure of the magnetization and susceptibility in low DC fields, the Neel temperature  value was determined from magnetization measurements taken in higher fields, where the FM-type contribution is fully saturated. These results are shown 
		in Fig.~\ref{figSM:M(H)};   panel 9c represents a typical temperature dependence of the saturation field $H_s$ and panels $(a, b)$,  for two field orientations, show the magnetization curves which  saturate above $H_s$ field. 
		
		\begin{figure}[h]
			\includegraphics[width=0.96\textwidth]{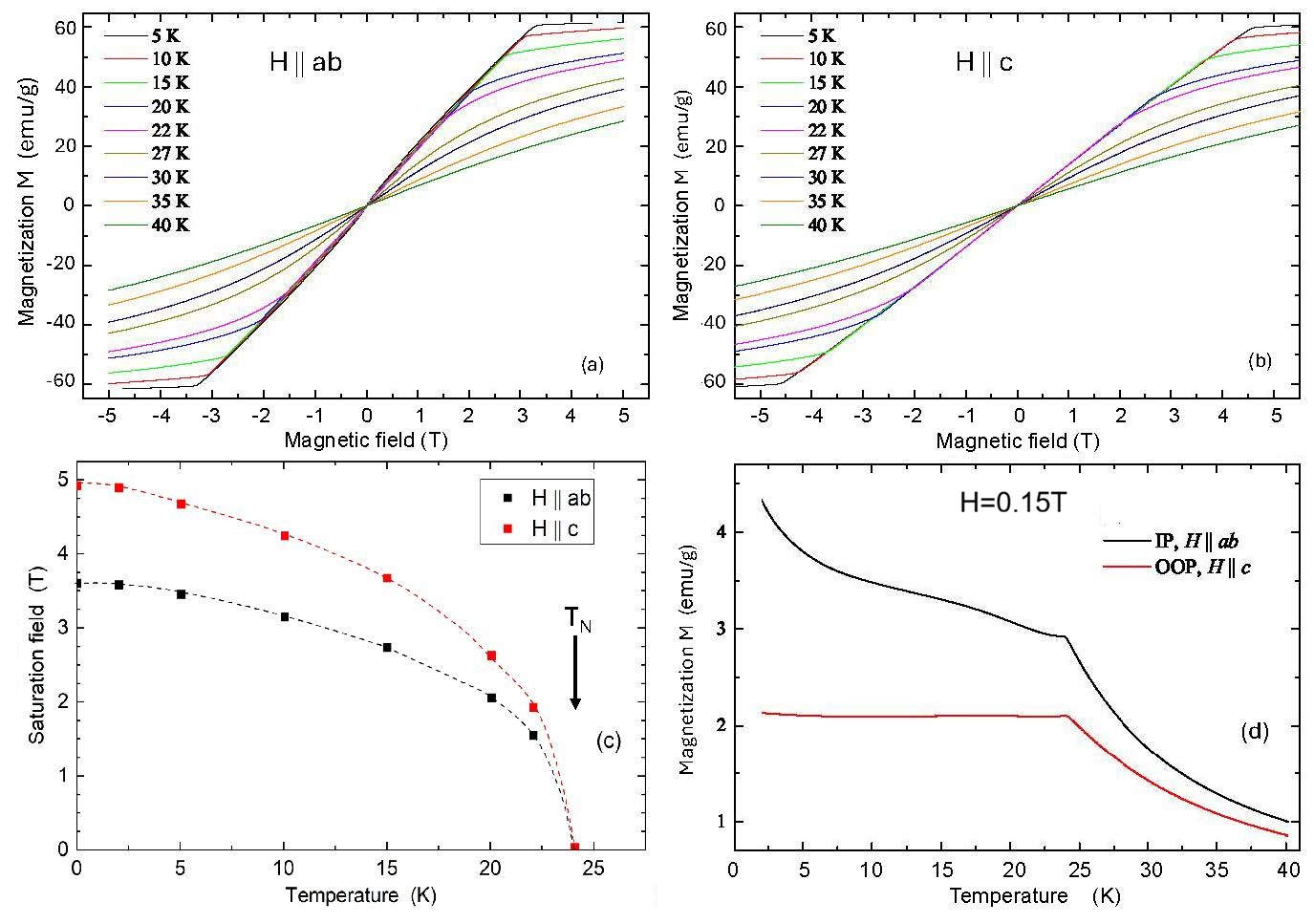}
			\caption{{\cb (a, b)Magnetization versus  magnetic field $M(H)$ at various temperatures in fields applied (a) in $()ab)$-plane,  and  (b) and out-of-plane, 	along the $c$ crystal direction. (c) The dependence of the saturation field on temperature for two orientations of the magnetic field showing definition of the Neel temperature $T_N$. (d) The dependence of magnetization on
					temperature $M(T )$ at  applied DC field $H= 0.15$ T for two orientations of the magnetic field. Adapted from Golovchanskyi et al. J. Magn. Magn. Mater., {\bf 562},	169713 (2022)  (Ref.~[12] of the main text).}
			}
			\label{figSM:M(H)}
		\end{figure}

		\newpage
		\section{Magnetic structure of the defect supercell: DFT results}
		The calculation of the magnetic structure was carried out using Wien2K software package.
		Figure \ref{figSM:magnetic structure}a shows the calculated  magnetic structure for the defect supercell
		in a collinear model in which the vectors are directed along the $c$-axis.
		We were unable to calculate such a complex structure in a non-collinear configuration where the vectors would be directed along $a$ or $b$.
		Various techniques were tried in VASP package, but it was not possible to bring the calculation to convergence (lack of convergence is 
		typical for non-collinear calculations due to the complexity of the problem).
		Therefore, a conventional collinear WIEN2k calculation with magnetic moments along $c$ was used as a model. This approximation provides an understanding of the influence of the presence of a defect in the structure. 
		In order to bring the calculated magnetic structure to the consistency with magnetization measurements (which unambigously show  that the magnetization vectors lie in the $ab$-plane) we present in Figure \ref{figSM:magnetic structure}c  
		a more realistic hypothesized magnetic structure in which  all vectors are rotated by $\pi/2$.
		
		The supercell shown in Fig.~\ref{figSM:magnetic structure}a contains 7 Eu atoms (considering  neighboring unit cells): each Eu atom
		in  positions 1  and  8  belongs to eight u.c.,  in positions $2, 4, 6,7$  - to one u.c., in position $5$ - to four u.c., 
		in position $3$ - the left atom belongs to one u.c., whereas two right ones  are located  out of the unit cell, and shown only as a guide to the eye, therefore they shouldn't be taken into account.
		The magnetic moments is counted similarly and  the total magnetic moment of the supercell  is associated with the magnetic moment of a single 
		Eu atom in position $2$ and equals to $6.70\mu_B$ (see also Section V of the main text).

		\begin{figure}[h]
			\includegraphics[width=0.48\textwidth]{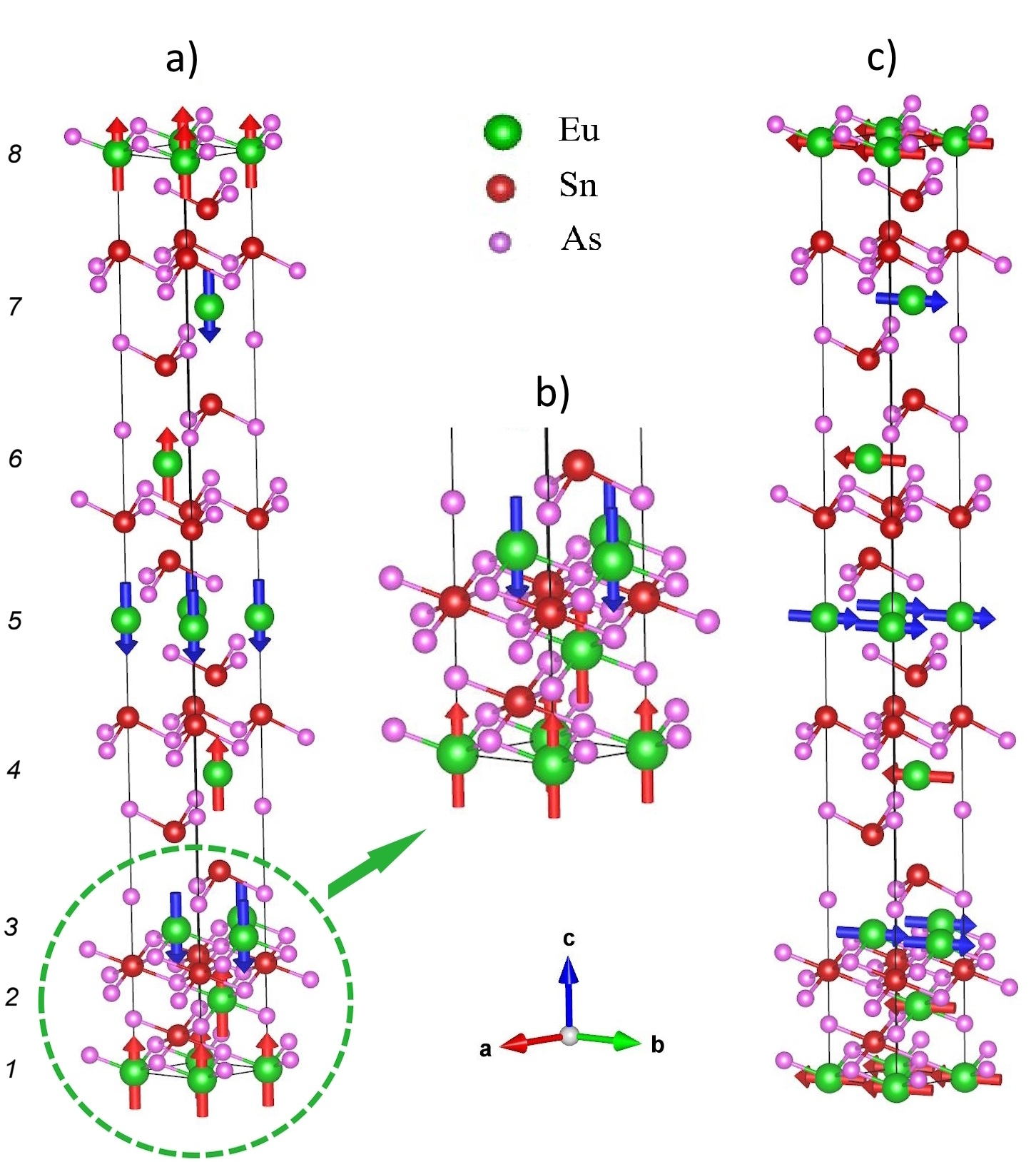}
			\caption{(a) Magnetic structure of the defect supercell Eu$_7$Sn$_{12}$As$_{14}$ calculated in the collinear configuration; (b) enlarged defect layer EuSnAs$_2$ of the supercell;  (c) hypothetical magnetic structure of the defect supercell Eu$_7$Sn$_{12}$As$_{14}$  for the easy-plane configuration.}
			\label{figSM:magnetic structure}
		\end{figure}
		
		\section{Anatomy of the magnetic ordering  of Eu moments in the host lattice and in the defect}
		It is tempting to discuss the physical mechanism underlying the ferromagnetic-type ordering of Eu in the planar defects.
		
		\underline{\bf Intraplane ordering}\\
		(i)	The ferromagnetic sign of the in-plane (IP) direct Eu-Eu exchange interaction in the host EuSn$_2$As$_2$ lattice is known from DFT calculations (Ref. [25] of the main text) with $J_1^{(IP)}= -1.98$meV. \\
		(ii)The IP distance between the neighbor Eu atoms is about the same in the  EuSn$_2$As$_2$ and EuSnAs$_2$ phases and, hence, provides the in-plane FM order for both. \vspace{0.1in}
		
		\underline{\bf Interplane ordering} is more complicated, since involves several competing channels of interplanar interaction. Moreover, the in-plane (IP) and out-of-plane (OP) interactions cannot be compared simply by contrasting the distances between nearest-neighbor Eu atoms, due to the spatial anisotropy of their 4f orbitals.
		Below, in two paragraphs, we overview the anatomy of the OOP interaction.\vspace{0.1in}
		
		\underline{EuSn$_2$As$_2$ (regular lattice)}:\\
		-Direct OOP exchange interaction for the regular ESA lattice with Eu-Eu interlayer distance of 1.3\,nm is $J_{\rm ex}^{\rm (OOP)} = -1.15$meV (of the FM-sign!) as calculated in ref. [25] of the main text.\\
		-Magnetic dipole OOP interaction is weak: for the two nearest Eu atoms separated by $d=1.3$nm, $J_d^{\rm (OOP)} \sim  -0.0015$meV (of the FM sign).\\
		-Spin-orbit interaction energy $\propto (v/c)^2(e^2/d)$ is also weak for the EuSn$_2$As$_2$  semimetal with small number of electrons, where $(v/c)^2  \sim 10^{-8}$. As a result, $J_{\rm SO}^{(\rm OOP)} \sim 0.016$ meV.\\
		-The RKKY interaction  is also small, $J_{\rm RKKY}^{\rm (OOP)} \sim -0.004$meV  and also has the FM sign (as estimated in Ref. [27] of the main text) for the given OOP distance and measured Fermi wave vector.\\
		-Double exchange interaction is missing in EuSn$_2$As$_2$  since Eu atoms are in a single valence state (see Fig.~5e of the main text).\\
		-Superexchange (the Kramers-Anderson mechanism) of the exchange coupling through the non-magnetic anions makes the largest contribution. The uppermost 4f-shell of Eu ions has 7 electrons and therefore is half-filled. Correspondingly, thanks to the Pauli exclusion principle and in accordance with the Goodenough–Kanamori rules, the interlayer coupling between two E$^{(2+)}$ magnetic ions with half-occupied orbitals, through an intermediary non-magnetic ions (Sn), is strong and has the antiferromagnetic sign. Its magnitude is found within the DFT calculations, while analytical evaluation is a task of quantum chemistry, beyond the scope of our paper.
		
		The net result  is $J^{\rm (OOP)} \sim +0.2$\,meV (of the AFM-sign). This value is consistent with the measured Neel temperature, 24K.\vspace{0.1in}
		
		\underline{Defect EuSnAs2 monolayer.}\\
		The in-plane distance between Eu atoms in the defect layer is about the same, as in the EuSn$_2$As$_2$  host lattice, 0.42\,nm, and therefore the IP exchange coupling should have ferromagnetic sign, similar to that in the regular EuSn$_2$As$_2$  lattice.
		
		The OOP coupling, however, is very different. 
		A reduction in the Eu–Eu interlayer distance from $d \approx 1.3$nm (regular EuSn$_2$As$_2$ structure) to $d \approx 0.8$nm (defect) leads to an enhancement of direct exchange, $J_{\rm ex}\propto  \exp(-d/2r_i)$, where the corresponding $r_i$ values lie in the range of $(0.05–0.109)$\,nm, with the first value representing the 4f-shell radius and the second - the ionic radius of Eu. The interlayer distances are large, $d/r_i \gg 1$, and, therefore, provide FM-sign of the direct exchange interaction (as clear, e.g.,  within the Bethe-Slater model). 
		By scaling $J_{\rm ex}^{\rm (OOP)}$ with $d$ we estimate $J_{\rm ex}^{(\rm OOP-defect)} \sim -(170-19)$\,meV, of the FM sign. 
		Furthermore, one can expect the primary opposing contribution (of the AFM sign) to the OOP  interaction—superexchange—to be smaller than in EuSn$_2$As$_2$ due to the absence of half of the Sn atoms.
		The large $J_{\rm ex}^{\rm (OOP-defect)}$ values exceed the contributions of all other OOP interaction  mechanisms, leading to the total interplane interaction in the planar defect to be ferromagnetic in nature. 
		
		As an additional result of our estimates,  we find the magnetic moment of Eu atom (in the defect EuSnAs$_2$ layer) in the supercell most likely should be directed parallel to its nearest neighbor, i.e. the lowest Eu atom $\#4$ in part B of the supercell (Fig.~5f of the main text). This is, of cause, confirmed by our DFT calculations.\vspace{0.1in}
		
		To summarize:\\ 
		- Similar enthalpy values and the variable valence of tin atoms facilitate the formation of the EuSnAs$_2$ phase as planar defects during the high-temperature synthesis of the target compound EuSn$_2$As$_2$.\\
		- Odd number of Eu-atoms in the defect layer paves the way for uncompensated magnetic moment. \\
		– This magnetic moment could potentially be compensated by all the other magnetic moments of Eu atoms in the supercell. However, the smaller interplane Eu-Eu distance and the absence of a Sn atom in the defect lead to the OOP exchange becoming ferromagnetic. 
		
		The above consideration is certainly of a qualitative nature. The exact results follow from our numerical DFT calculations, where all atomic orbitals and interaction channels are included. Exact analytical consideration of the magnetic ordering is beyond the scope of our paper.
	}
	
\end{document}